\documentclass[12pt]{article}

\pdfoutput=1

\usepackage[parfill]{parskip}
\usepackage[table]{xcolor}
\usepackage{amsthm,enumerate}
\usepackage{mathtools}
\usepackage[normalem]{ulem}
\usepackage[T1]{fontenc}
\usepackage[utf8]{inputenc}
\usepackage[thinc]{esdiff}
\usepackage{color}
\usepackage{fancyhdr}
\usepackage{calc}
\usepackage{url}
\usepackage{xcolor}
\usepackage{xspace}

\usepackage[all,cmtip]{xy} 
\usepackage{tikz-cd}

\usepackage[top=80pt,bottom=85pt,left=85pt,right=85pt]{geometry}
\usepackage{amsmath,amssymb,graphicx,float,color,tikz,cite,savesym,wasysym,subfigure,wrapfig}
\usepackage{caption}
\usepackage[debug,pageanchor=false]{hyperref}
\definecolor{link}{rgb}{1,0.45,0.05}
\hypersetup{colorlinks=true,linkcolor=link,citecolor=link,urlcolor=link,linktocpage}

\newcommand{\R}{\mathbb{R}}

\newcommand{\dd}{\mathrm{d}}

\newcommand{\rj}{r_\text{join}}
\newcommand{\TT}{\mathbb{T}}
\newcommand{\ZZ}{\mathbb{Z}}
\newcommand{\RR}{\mathbb{R}}

\newcommand{\EE}{\mathsf{E}}
\newcommand{\BB}{\mathsf{B}}

\usepackage{booktabs}

\theoremstyle{plain}

\newtheorem*{theorem*}{Theorem}
\newtheorem*{maintheorem*}{Main Theorem}

\theoremstyle{definition}

\newtheorem*{definition*}{Definition}
\newtheorem*{remark*}{Remark}

\makeatletter
\@addtoreset{equation}{section}
\makeatother

\begin{document}

\begin{titlepage}

	\begin{center}

	\vskip .5in 
	\noindent

	{\Large \bf{Machine Learning Gravity Compactifications\\on Negatively Curved Manifolds}}

	\bigskip\medskip
	 G. Bruno De Luca\\

	\bigskip\medskip
	{\small 
$^1$ Stanford Institute for Theoretical Physics, Stanford University,\\
382 Via Pueblo Mall, Stanford, CA 94305, United States
}

   \vskip .5cm 
	{\small \tt gbdeluca@stanford.edu}
	\vskip .9cm 
	     	{\bf Abstract }
	\vskip .1in
	\end{center}

	\noindent
	 {Constructing the landscape of vacua of higher-dimensional theories of gravity by directly solving the low-energy (semi-)classical equations of motion is notoriously difficult. In this work, we investigate the feasibility of Machine Learning techniques as tools for solving the equations of motion for general warped gravity compactifications.  As a proof-of-concept we use Neural Networks to solve the Einstein PDEs on non-trivial three manifolds obtained by filling one or more cusps of hyperbolic manifolds. While in three dimensions an Einstein metric is also locally hyperbolic, the generality and scalability of Machine Learning methods, the availability of explicit families of hyperbolic manifolds in higher dimensions, and the universality of the filling procedure strongly suggest that the methods and code developed in this work can be of broader applicability. Specifically, they can be used to tackle both the geometric problem of numerically constructing novel higher-dimensional negatively curved Einstein metrics, as well as the physical problem of constructing four-dimensional de Sitter compactifications of M-theory on the same manifolds.}
	
	\vfill
	\eject

	\end{titlepage}

\tableofcontents
\newpage
\section{Introduction}

In theories with extra dimensions, four-dimensional vacua are non-trivial even classically. In fact, even a configuration that looks like a vacuum in four dimensions can have (and often has) a very rich structure for the geometry and the matter fields in the extra dimensions.
The study of these structures, which are simultaneously rich and highly constrained by the UV completion of the theory, is important to extract the four-dimensional physics, as well as for holographic approaches to quantum gravity.

As we will discuss in detail below, the problem of directly solving the equations of motion for vacuum compactifications is computationally challenging, and a popular approach to bypass this challenge is to exploit supersymmetry in some form.  This is the case, for example, of the famous KKLT proposal \cite{Kachru:2003aw} for obtaining de Sitter vacua, which uses as starting point a supersymmetric Calabi-Yau compactification, on top of which supersymmetry-breaking effects are added. But this is true also for AdS: most of the explicitly known AdS compactifications of string/M-theory are either supersymmetric, obtained by starting from supersymmetric compactifications and turning on supersymmetry breaking effects, or as non-supersymmetric vacua of lower-dimensional supergravities obtained from reduction around supersymmetry vacua. Supersymmetry allows to control certain higher-order corrections and instabilities, and at the technical level it simplifies the solution of the classical equations of motion by reducing the problem of solving non-linear second order PDEs to the simpler problem of enforcing the first order supersymmetry equations\footnote{See \cite{legramandi2020breaking} for a method to deform the first order system of BPS equations in a way that breaks supersymmetry but still implies the equations of motion, applicable to certain classes of starting supersymmetric configurations.} (see e.g.~the textbook \cite{Tomasiello:2022dwe} for an introduction to these methods and constructions). 

But supersymmetry is not compatible with a universe in a de Sitter phase, and it is also believed to be an obstacle for obtaining four-dimensional AdS vacua with small internal spaces \cite{Polchinski:2009ch}, with proposed examples having little ($\mathcal{N}= 1$) \cite{DeWolfe:2005uu, Polchinski:2009ch, Cribiori:2021djm, Kachru:2003aw, Demirtas:2021nlu} or zero \cite{DeLuca:2022inb} supersymmetry.\footnote{See also \cite{Montero:2024qtz} for a recent analysis hinting to the possibility that in the scale-separated DGKT vacuum $\mathcal{N} = 1$ supersymmetry might be broken by quantum effects.} In addition, when looking for de Sitter vacua, the constraints imposed by lower-dimensional supersymmetry might actually make the problem unnecessarily harder by restricting the positive contributions to the four-dimensional potential \cite{Flauger:2022hie}, forcing one to pivot, for example,  to non-perturbative effects \cite{Kachru:2003aw}. 

Given the importance of non-supersymmetric compactifications to describe realistic four-dimensional physics, and that the vast majority of possible configurations are not supersymmetric \cite{Silverstein:2016ggb}, it is important to explore the landscape of possible compactifications relying as little as possible on supersymmetry, avoiding lamppost effects.
In fact, the technical and computational challenges in the construction of non-supersymmetric configurations should not be mistaken for roadblocks in extracting non-supersymmetric physics, as standard EFT methods, perturbative approaches, and concrete approximation schemes, are very powerful when under control and have led to rich phenomenology and studies of de Sitter, inflationary, and more general non-supersymmetric compactifications (see e.g.~the reviews \cite{Flauger:2022hie, Silverstein:2016ggb, McAllister:2023vgy}).  Moreover, in appropriate cohomogeneity-one setups, one can sometimes reduce the equations of motion to a systems of ODEs, which are readily solved with standard numerical methods, allowing to construct de Sitter \cite{strings-talk, Cordova:2018dbb, Cordova:2019cvf, Burgess:2024jkx} and non-supersymmetric AdS (e.g.~\cite{Cordova:2018eba}) compactifications of ten-dimensional supergravity theories.

Nonetheless, our ability to extract detailed features of a given vacuum and of understanding general properties of non-supersymmetric compactifications would be enhanced by the availability of efficient methods to construct general gravity compactifications by a direct analysis of the (semi-) classical equations of motion. As we will review in Sec.~\ref{sec:eoms}, such a method should be able to solve Einstein-like PDEs on non-trivial manifolds in relatively high number of dimensions (i.e.~6 or 7 for compactifications of string/M-theory in common geometric phases). 

This seems like a daunting task, and indeed, unfortunately, standard numerical approaches struggle particularly at high dimensionality. As a consequence, direct solutions of the Einstein-like PDEs in compactifications are scarce. Some notable examples are the construction of Kaluza-Klein black holes \cite{Headrick:2009pv} or the analysis of related geometric problems, such as the construction of Calabi-Yau metrics \cite{Headrick:2005ch,Douglas:2006rr, Headrick:2009jz} or K\"ahler-Einstein metrics \cite{Doran:2007zn}, where the Einstein PDEs can be reduced to more tractable scalar PDEs.

Luckily in recent years Machine Learning has emerged as a powerful tool to tackle a variety of computational problems. While traditionally developed for solving data-intensive problems, it can also be applied to situations with little or no data. One very general framework, sometimes named \emph{physics informed machine learning} \cite{PIML} has emerged for applying these methods to the solution of PDEs, starting with the pioneering work \cite{lagaris1998artificial}. Outside of the PDE domain, similar techniques have seen an explosion of applications to high energy physics, most recently, for example, in bootstrapping string amplitudes \cite{Bhat:2024agd}, in learning Seiberg dualities \cite{Capuozzo:2024vdw}, and in the study of nonperturbative scattering amplitudes \cite{Gumus:2024lmj}.\footnote{See also \cite{Gukov:2024buj} for a recent discussion on techniques to obtain exact results in theoretical physics from ML methods and \cite{Liu:2024swq} for a novel architecture that can help with this task.}

While Machine Learning methods for solving PDEs lack the same convergence guarantees of more traditional methods, they have been found to work well in very high number of dimensions \cite{sirignano2018dgm}, a regime where classical methods struggle.  In the context of string compactifications, Machine Learning techniques have been successfully applied to the problem of constructing Calabi-Yau metrics \cite{Ashmore:2019wzb, Douglas:2020hpv, Jejjala:2020wcc, Larfors:2021pbb, Larfors:2022nep, Gerdes:2022nzr} and their corrections \cite{Fraser-Taliente:2024etl}. In these cases the task is simplified by the use of analytic properties of Calabi-Yau geometry. By exploiting these properties, one does not have to directly enforce the vanishing of the Ricci \emph{tensor}, greatly simplifying the computational cost, at the price of general applicability (see e.g.~the pedagogical review \cite{Anderson:2023viv}).

In this work, we start an exploration of the viability of Machine Learning methods to construct \emph{general warped compactifications}, without relying on particular symmetric or analytic structures, by directly solving the Einstein-like PDEs, reviewed in Sec.~\ref{sec:eoms}, on patchwise defined compactification manifolds, extending earlier work in \cite[Sec.~8]{deluca-silverstein-torroba}. An immediate advantage of such an approach would be its ready applicability to a variety of compactifications, directly allowing for the inclusion of matter fields and their backreaction in the equations of motion. 

As we will discuss in Sec.~\ref{sub:mlgeo}, in the Machine Learning approach the problem of solving the Einstein-like compactification equations is translated into an optimization problem in the parameter space of suitably defined Neural Networks. While solving this in general remains a very hard problem, the Machine Learning formulation has many practical advantages that empirically suggest its viability as a general-purpose method for constructing general gravity compactifications.  In particular, this formulation allows to take advantage of the recent progress in the development of optimization algorithms for very high-dimensional landscapes, such as~Adam \cite{kingma2014adam} and its relatives, ECD algorithms based on chaotic Hamiltonian dynamics\footnote{These energy-conserving Hamiltonian methods have found application to more general physics and scientific problems. For example, the analytic handles provided by the energy conserving Hamiltonian dynamics are being used in \cite{wip-LHC} for precision optimization for inference in particle physics. \cite{MCHMC} developed gradient-based sampling methods based on the ECD dynamics, which have been successfully applied to proof-of-concepts analyses of CMB data in \cite{Bonici:2023xjk} and lattice field theory in \cite{Robnik:2023pgt}.} \cite{BBI, ECDSep}, or recent higher order methods \cite{pagliardini2024ademamix}. Another practical advantage is the possibility to directly benefit from fast progress in software (automatic differentiation packages and well-developed ML libraries) and hardware (such as GPUs and TPUs) tailored for Machine Learning applications.

In addition, as we will discuss in Sec.~\ref{sub:method},  the ML formulation allows for a simple flexible way to encode prior knowledge of \emph{approximate solutions}, which can often be constructed with more traditional methods, providing a general way to incorporate previous knowledge of the target solution. 

In Sec.~\ref{sec:MLE}, we apply this formulation and these methods to a geometric problem of direct relevance for the construction of smooth four-dimensional de Sitter compactifications  of M-theory, as introduced in \cite{deluca-silverstein-torroba} and briefly reviewed in Sec.~\ref{sub:dshyp}. Specifically, we discuss the construction of smooth Einstein metrics corresponding to the filling of cusped hyperbolic manifolds \cite{anderson-dehn,bamler2012construction}, which we review in Sec.~\ref{sub:filling}. As a proof-of-concept, we study in detail a non-diagonal filling of a single cusp of the three-dimensional hyperbolic manifold $M_3$, a particular example of an explicit family of hyperbolic manifolds in dimension $3\leqslant n \leqslant 8$ constructed in \cite{italiano2024hyperbolic} and reviewed in Sec.~\ref{sub:M3}. While in three dimensions an Einstein metric is also locally hyperbolic, the filling procedure, the explicit family of manifolds, as well as our approach of directly solving the Einstein equations, are readily generalizable to higher dimensionality. These features make our proof-of-concept study directly scalable to the construction of novel Einstein metrics in higher dimensions, and serve as a basis for the addition of the other physical fields in addition to the metric, building towards a method suitable for solving the full set equations of motion for general gravity compactifications on non-trivial manifolds. We collect the details about our numerical procedure and results in Sec.~\ref{sub:ML3d}. 

We conclude in Sec.~\ref{sec:concl} with a brief summary and discussion.

The code to reproduce and validate our results is freely available at \href{https://github.com/gbdl/fillings}{github.com/gbdl/fillings} under a GNU GPL license \cite{gplv3}.

\section{Machine Learning methods for gravity compactifications}
We start in Sec.~\ref{sec:eoms} by reviewing the equations of motion for general (semi-)classical gravity compactifications, which are composed by a system of non-linear second order PDEs. In Sec.~\ref{sub:mlgeo} we then briefly review how Machine Learning methods can be designed to tackle PDE problems in high dimensions, and we present in  Sec.~\ref{sub:method} a step-by-step procedure tailored to these geometrical equations of motion. We conclude in Sec.~\ref{sub:dshyp} by reviewing an example of a natural setup where these methods can be applied: the problem of finding four-dimensional compactifications of M-theory on negatively curved spaces.

\subsection{The equations of motion}\label{sec:eoms}
We are working with theories that at low energies reduce to $D$-dimensional Einstein gravity, with a standard kinetic term, plus some matter content.
That is, our starting point is a $D$-dimensional gravitational theory with an
action of the form
\begin{equation}
	S_D = m_D^{D - 2} \int \sqrt{- g_D} R_D + S_{\text{matter}} +
	\ldots \label{eq:genAction}
\end{equation}
where $m_D$ is the $D$-dimensional Planck mass, and the dots represent corrections to the low-energy action coming from the UV completion.  $S_\text{matter}$ is the action for the matter fields in the theory.
Notable examples of such theories are the low-energy limits of string/M-theory, but the discussion in this section is more general.

We are interested in lower $d$-dimensional vacua of these theories, and specifically in the case $d = 4$, to which we will specialize momentarily. By definition, vacua are configurations with maximal $d$-dimensional symmetry, implying for the higher-dimensional space-time a decomposition of the form
\begin{equation}
	\text{ds}^2_D (x, y) \equiv e^{2 A (y)}\left( \text{ds}^2_{d, \Lambda} (x) +
	\text{ds}^2_n (y) \right) \, . \label{eq:metAns}
\end{equation}
We use coordinates $x$ with indices $\mu, \nu, \ldots$ to denote the $d$-dimensional space-time, and coordinates $y$ with indices and $m, n,\ldots$ to denote indices in the extra $n$ dimensions. A similar symmetry-preserving decomposition applies in the vacuum to the other fields of the theory.	

There are various ways to organize the equations of motion obtained from \eqref{eq:genAction} and specialized to the ansatz \eqref{eq:metAns}.  These equations are composed by a scalar equation, corresponding to the higher-dimensional Einstein equation along the vacuum directions, plus all the Einstein equations along the internal directions. A natural way to organize them for our discussion is as in \cite[Sec.~2.1]{DeLuca:2022inb}
\begin{align}
	&\left[\Delta +(D-2)\left(\Lambda-\frac{1}{d}\hat T ^{(d)} \right)\right] e^f = 0\label{eq:Schr}\\
	&\text{Ric}^f_{mn} = \Lambda g_{mn} + \tilde{T}_{mn}\label{eq:EinstInt}
\end{align}
where $f \equiv (D-2)A$ and the `warped' Ricci tensor\footnote{This and similar combinations of Ricci tensor and derivatives of the warping naturally appear in the context of optimal transport theory and Bakry-\'Emery geometry, where they are shown to control the spectrum of certain warped Laplacians. These results translate into general theorems constraining the universal spin 2 Kaluza-Klein spectrum for general warped compactifications \cite{DeLuca:2021mcj,DeLuca:2021ojx,DeLuca:2022inb}. } $\text{Ric}^f$ is defined as
\begin{equation}
	\text{Ric}^f_{mn}\equiv	R_{mn} - \nabla_m \nabla_n f + \frac{1}{D-2} \nabla_m f \nabla_n f \;.
\end{equation}
The combinations of stress-energy tensors appearing in these equations are defined as
\begin{equation}
	\hat{T}^{(d)} \equiv g^{(d)\mu\nu}\hat{T}_{\mu\nu}, \quad \tilde{T}_{mn} \equiv \frac{1}{2}m_D^{2-D}\left(T_{mn}^{(D)} - \frac{1}{d}g_{mn}T^{(d)}\right), \hat{T}_{MN} \equiv \frac{1}{2}\kappa^2\left(T_{MN} - g_{MN}\frac{T}{D-2}\right) .
\end{equation}
In addition, the equations of motion of the theory include the equations of motion for the matter fields. 

In presence of non-trivial warping $e^f$, the equations of motion for general vacuum compactifications are organized into a Schr\"odinger-like equation \eqref{eq:Schr} for the warping\footnote{Not to be confused with the inhomogeneous Schr\"odinger equation that defines the $4d$ effective potential as derived in \cite{Douglas:2009zn}.} and an Einstein-like equation \eqref{eq:EinstInt} for the warped Ricci tensor.

A well-known constraint on allowed compactifications  can be obtained by averaging the Schr\"odinger equation \eqref{eq:Schr} over the internal space, \emph{under the assumption of a smooth internal space without boundaries}. In this case,  the integral of the stress-energy tensor combination $\hat T ^{(d)}$ determines the cosmological constant $\Lambda$. In particular, restricting only to classical sources and localized objects with positive tension, this integral is negative, forbidding the existence of de Sitter solutions in this regime \cite{gibbons-nogo,maldacena-nunez}. 

Without this restrictive assumption, both signs for the cosmological constant are possible. 
Terms that contribute positively to the cosmological constant can be classical but stringy, such as localized sources with negative tension, quantum, such as Casimir energy, or also non-perturbative or higher-derivative effects. 
The cosmological constant is then determined by the cancellation between positive and negative contributions to the stress-energy tensor combination in \eqref{eq:Schr}. 

In Sec.~\ref{sub:dshyp} we review a proposal for dS$_4$ vacua in M-theory where the cancellation is between classical effects in the form of fluxes and quantum effects in the form of Casimir energies.

Having found a situation where $\Lambda >0$ is consistent with the equations motion, the problem of finding explicit solutions for the physical fields is highly challenging, even numerically. 
As opposed to $\Lambda < 0$, in this case we cannot look for supersymmetric configurations, where it is possible to trade the second order system of equations of motion \eqref{eq:Schr} and \eqref{eq:EinstInt} for a simpler system of first order equations that impose supersymmetry and implies the equations of motion (see e.g.~the textbook \cite{Tomasiello:2022dwe}), and we thus have to work with this full set of non-linear geometrical PDEs in relatively high number of dimensions.

When the goal is to extract physical properties of a class of compactifications, this computational obstacle is not necessarily a roadblock: manually-constructed piecewise approximate solution to the full equations can provide enough information, provided these are obtained in a regime where the corrections are small.
Nevertheless, a natural question is whether it is possible to develop a general method flexible enough to produce detailed numerical solutions of these geometrical equations, a problem to which we will now turn.

\subsection{Machine Learning methods for geometrical PDEs}\label{sub:mlgeo}
Having identified a semi-classical setup in which the balancing of stress-energy sources is consistent with the sign of the cosmological constant, finding an explicit detailed solution requires to solve the equations of motions \eqref{eq:Schr}, \eqref{eq:EinstInt}.

As opposed to methods that reduce the Einstein equations to simpler scalar PDEs exploiting geometric or analytic properties of the compactification (e.g.~the use of the Monge-Amp\`ere equation for Calabi-Yau compactifications \cite{Anderson:2023viv}), the focus of this work is to discuss methods to directly study the system \eqref{eq:Schr}, \eqref{eq:EinstInt}.

The general system of equations is composed by an elliptic PDE \eqref{eq:Schr}, a weakly-elliptic Riemannian Einstein-like equation \eqref{eq:EinstInt}, plus usually elliptic equations of motion for static matter field configurations in the internal space.  The weak form of ellipticity of \eqref{eq:EinstInt} is a consequence of invariance by local diffeomorphisms, which manifests itself in  \eqref{eq:EinstInt} being an underdetermined system for $g_{mn}$. Specifically, out of the $n(n+1)/2$ metric components, the Einstein equations only fix $n(n-1)/2$, since $n$ equations are redundant as a consequence of the Bianchi identity. We refer the reader to \cite[Sec.~2]{Headrick:2009pv} for a detailed discussion of these features.
A natural way to fix the gauge is to select and fix a reference vector field $\xi$, thus breaking gauge invariance and providing the extra $n$ equations needed.  Given a reference metric $\bar{g}$, this can be constructed as $\xi^m \equiv g^{n p} \left( \Gamma_{n p}^m -\bar{\Gamma}_{n p}^m\right)$, and the gauge is fixed by setting $\xi^m = 0$. This condition can be elegantly imposed simultaneously while solving the Einstein equations by redefining the Ricci tensor as $R_{mn} \to R_{mn} +\nabla_{(m} \xi_{n)}$, resulting in the so-called Einstein-De Turck equations \cite{Headrick:2009pv}.  
As we will now discuss, when using Machine Learning methods one could either follow this approach or directly enforce $\xi = 0$ in the loss function. We also note here that the warping equation \eqref{eq:Schr} and the Einstein-like equations \eqref{eq:EinstInt} are coupled, and a more detailed analysis could take their coupling into account.

Having fixed the gauge, we have a well-posed system, which is still extremely challenging to solve numerically with standard methods. The main obstacle stems from the fact that we are seeking to solve a non-linear system of PDEs in a relatively high number of dimensions, typically six or seven for string/M-theory compactifications. In this scenario, standard methods such as finite differences, finite elements, or finite volumes suffer from the \emph{curse of dimensionality}, an exponential increase in computational complexity with the number of dimensions. 

Depending on the specificity of the problem and other constraints, refined methods can be applied to try to beat the curse of dimensionality, from spectral methods to multigrid approaches. In this work, we will focus instead on Machine Learning methods for PDEs, which have attracted interest due their versatility, ease of implementation, and ability to leverage modern hardware and algorithms designed for more general Machine Learning problems.
The literature on this topic is vast, and we will review and extend in Sec.~\ref{sub:method} the simplest approach with the goal of providing a self-contained description of the methods we will apply to a specific example in Sec.~\ref{sub:ML3d}. We also refer the reader to \cite[Sec.~8]{deluca-silverstein-torroba} for a review with a focus on compactifications and to \cite{wang2023expert} for a recent review on various techniques to improve PDE solving via ML. 

A natural starting point in applying Machine Learning to solve PDEs, as originally explored in \cite{lagaris1998artificial}, is to use a Neural Network as an ansatz for the solution. A Neural Network can simply be understood as a parameterized function $\mathcal{N}(x; \theta)$, and the parameters $\theta$ are adjusted to a value $\theta^\star$ such that $\mathcal{N}(x; \theta^\star)$ approximates as close as possible a solution to the PDEs, including boundary and consistency conditions and other constraints. To find these parameters a common approach is to design a \emph{loss function} $\mathcal{L}$, a function in parameter space minimized at $\theta = \theta^\star$. 

From a physics point of view, rephrasing the problem in this way translates it into a dynamical systems problem: to solve the PDE we need to prescribe, and follow, a dynamics in parameter space that explores the loss landscape $\mathcal{L}$ and equilibrates at $\theta = \theta^\star$, possibly in a short amount of time. This rephrasing applies to general machine learning problems. 

An efficient approach to prescribe such a dynamics is to take the neural network and the loss functions to be differentiable functions, so that we can explore the loss landscape via gradient-based methods. The standard approach uses gradient descent-like dynamics (or \emph{optimizers}), such as `gradient descent with momentum' \cite{SGDM-1, SGDM-2} and more refined versions such as Adam \cite{kingma2014adam}. Physically they correspond to friction-dominated motion on the loss landscape, which generically converges to local minima.  A natural question from the physics perspective is whether there are other physical dynamics with the property of converging to $\theta = \theta^\star$. Along these lines, \cite{BBI, ECDSep} develops alternative optimizers that employ instead energy-conserving Hamiltonian dynamics in a chaotic regime, where convergence on a desired region of the loss landscape is ensured by localizing the Liouville measure on $\theta = \theta^\star$ via different physical mechanisms, such as relativistic dynamics. In our experiments in Sec.~\ref{sec:res} we will employ an optimization method in this class.

In the next section, we specialize these ideas to geometrical PDEs. 

\subsubsection{A method for compactifications}\label{sub:method}

To set up the notation, we assume we have set of differential geometric equations $\{\EE_a\}$, $a = 1,\dots, n_{\mathsf{E}}$,  for a set of tensor fields $\{\varphi_i\}$, $i = 1, \dots, n_{\varphi}$,  on a manifold $M$, that is, we have to solve
\begin{equation}
	\mathsf{E}_a(\{\varphi_i(x)\}, x) = 0\;,\qquad \qquad \forall a \in [1, n_\EE]\;,  x \in M. 
\end{equation}
For gravity compactifications, $\{\EE_a\}$ is composed by the Schr\"odinger equation \eqref{eq:Schr}, the Einstein-like equations \eqref{eq:EinstInt} plus the equations of motion for matter fields. 

For concreteness and for a direct application to the example studied in Sec.~\ref{sub:ML3d}, we assume in this section that we can parameterize $M$ as a single region $\Omega \subset \RR^n$, with appropriately specified boundary conditions at its boundary $\partial \Omega$. Specifically, we denote the equations that impose the boundary conditions by $\BB_q$, $q = 1, \dots, n_{\BB}$, such that we have to impose
\begin{equation}
	\BB_q(\{\varphi_i(\tilde x)\}, \tilde x) = 0 \qquad\qquad \forall q \in [1, n_\BB]\;,  \tilde{x} \in \partial \Omega.
\end{equation}
In the example developed in Sec.~\ref{sec:MLE}, $\{\BB_q\}$ will include conditions that enforce the matching of the induced metric (continuity), and of the extrinsic curvature (differentiability), on identified regions of $\partial \Omega$. 
The general case with $M$ decomposed in multiple regions $\{\Omega_\mu\}$, with boundary conditions gluing different parts of $\{\partial \Omega_\mu\}$, follows as a straight-forward generalization of our discussion below. 

Finally, we might also want to impose constraints, such as for example the gauge fixing $\xi = 0$ described at the beginning of Sec.~\ref{sub:mlgeo}. We denote these as $\mathsf{C}_s$, $s \in 1, \dots, n_\mathsf{C}$, and impose 
\begin{equation}
	\mathsf{C}_s(\{\varphi_i(x)\}, x) = 0 \qquad\qquad \forall s \in [1,n_\mathsf{C}]\;,   x \in \Omega.
\end{equation}
In case $M$ is decomposed in multiple $\Omega_\mu$, the constraints can include the matching of the transitions function in the overlaps of the $\{\Omega_\mu\}$, depending on the details of the decomposition.

Optionally, we also assume that we have an \emph{approximate solution} $\{\bar{\varphi}_i\}$, which we can use as a starting point. While this is not strictly needed if the dynamics avoids local minima where $\mathcal L > 0$, having a good starting point can greatly accelerate convergence to the desired solution. 

Having set the stage, we can perform the following steps. 
\begin{enumerate}
	\item[] \textbf{Step 1}:  Parametrize the tensor fields $\{\varphi_i\}$ as a set of neural networks $\{\mathcal{N}_p(x; \theta)\}$. For example, if the tensor field is a metric on an $n$-dimensional manifold $M_n$, in the local patch $\Omega$ we can parameterize its $n(n-1)/2$ independent components as different  $\mathcal{N}_p(x; \theta): \Omega\to \RR$, or as a single neural network   $\mathcal{N}(x; \theta): \Omega\to \RR^{n(n-1)/2}$. As a result, we can write $\{\varphi_i(x)\}_\theta$, with $\theta$ representing collectively the parameters of $\{\mathcal{N}_p\}$.
	
	\item[] \textbf{Step 2}: Choose a sampling scheme. During the training we will need to sample points $x\in  \Omega$ and $\tilde{x} \in \partial \Omega$, and we define probability distributions $\rho$ and $\tilde{\rho}$ to perform this sampling. Common choices for these probability distributions are either uniform in $\Omega$, or, more geometrically, uniform according to a natural geometric measure on $M_n$, such as the Riemannian volume form of the starting approximate solution for the metric. In addition, $\rho$ and $\tilde{\rho}$ can be time-dependent, i.e.~changing during the training, for example by increasing the mass in regions where the error is larger. We will introduce a simple geometric and time-dependent scheme in Sec.~\ref{sub:ML3d}, and we refer the reader to \cite{wu2023comprehensive} for a review of different sampling strategies.

	\item[] \textbf{Step 3} (optional): Pre-train the neural networks to match the approximate solutions $\{\bar\varphi_i\}$. This can be set up as a simple \emph{supervised learning} problem, where the parameters $\theta$ are found by minimizing a loss function of the form
	\begin{equation}
		\mathcal{L}_{\text{pre}}(\theta) \equiv \sum_{i = 1}^{n_\varphi} \left(\sum_{x \sim \rho}  \left| \varphi_i(x, \theta) - \bar \varphi_i(x)\right|^2 + \sum_{\tilde x \sim \tilde\rho} \left| \varphi_i(\tilde x; \theta) - \bar \varphi_i(\tilde x)\right|^2\right) \;.
	\end{equation}
	This supervised learning problem is generally relatively easy to solve with a variety of standard methods, however, imposing extra constraints, such as explicit matching of first and second derivatives, can make it computationally more challenging, as we will see in detail in Sec.~\ref{sub:ML3d}.\footnote{
	Alternatively, a popular choice is to impose that $\varphi$ starts out as $\bar \varphi$ by an explicit parameterization (e.g.~additive or multiplicative). See for example \cite[Table 1]{Anderson:2023viv} for a review of this approach for construction of numerical Calabi-Yau metrics. While this avoids the pre-training step, it requires a manual parameterization of the solution in terms of the approximate one. In the example studied in Sec.~\ref{sec:MLE}, we found the pre-training approach to be more flexible and faster in adapting from the approximate solution to the final one for our class of problems.}

	\item[] \textbf{Step 4}: Construct a loss function for the full problem. At this step, $\{\varphi_i(x) \}_\theta$ represent an approximate solution to the equations $\EE_a$, the boundary conditions $\BB_q$ and the constraints $\mathsf{C}_s$. We then construct a loss function that is minimized when these equations are solved exactly. The simplest choice is \begin{equation}\label{eq:loss}
		\begin{split}			
		\mathcal{L}(\theta) &\equiv \sum_{x\sim \rho} \left(\sum_{a=1}^{n_\EE} \gamma_{\EE} \Big{|}\Big{|}\EE_a(\{\varphi_i(x)\}_\theta, x)\Big{|}\Big{|} + \sum_{s=1}^{n_\mathsf{C}} \gamma_{\mathsf{C}}\Big{|}\Big{|}\mathsf{C}_a(\{\varphi_i(x)\}_\theta, x)\Big{|}\Big{|}  \right) + \\
		 &+  \sum_{\tilde x\sim \tilde \rho}\sum_{q=1}^{n_\BB} \gamma_\BB \Big{|}\Big{|}\BB_q(\{\varphi_i(\tilde x)\}_\theta, \tilde x)\Big{|}\Big{|}\;,
	\end{split}
	\end{equation} 
	where $\gamma_{\EE}$, $\gamma_{\mathsf{C}}$, and $\gamma_\BB$ are \emph{balancing coefficients}, which can be either kept fixed or dynamically adjusted during the minimization of  $\mathcal{L}$ (see e.g.~\cite[Sec.~5.2]{wang2023expert} for a discussion of different balancing strategies). With the symbol $||\cdot ||$ we denote an appropriate norm for the corresponding tensor structure, which can be chosen depending on the problem. 

	\item[] \textbf{Step 5} (training): Choose a dynamics and follow it. 
	We minimize $\mathcal{L}(\theta)$ in \eqref{eq:loss} for $\theta$, starting from the initial condition $\theta = \theta_0$, with $\theta_0$ determined after the pre-training phase, or as a random initial condition if the optional pre-training step is not carried out. 
	As discussed above, this problem can be solved by choosing an appropriate (discrete) dynamics that at long times converges to $\theta^\star$ where $\mathcal{L}(\theta^\star) \approx 0$. Notice that in this problem we are particularly interested in global minima, as local minima would generically imply that the equations are not solved, suggesting a natural application for dynamics that cannot stop at local minima due to conservation laws, such as the ECD optimizers introduced in \cite{BBI,ECDSep}. On the other hand, stochastic effects, together with the expectation that at large enough dimensionality high local minima might be disfavored \cite{bahri2020statistical}, could help more standard optimization methods such as Stochastic Gradient Descent with Momentum\cite{SGDM-1,SGDM-2} or Adam \cite{kingma2014adam}. In Sec.~\ref{sec:res} we test the $\mathsf{ECDSep}$ optimizer on the three-dimensional problem studied in Sec.~\ref{sec:MLE}, leaving a detailed comparison on higher-dimensional examples to future work.
\end{enumerate}

We conclude with a few comments. 
Following a standard approach, we sample a set of points from $\rho$ and $\tilde \rho$ at each step of the optimization procedure, to avoid overfitting (i.e. the risk of finding $\{\varphi_i\}$ that solve the equations only at the sampled point, but do poorly elsewhere). This feature makes the algorithm more scalable in high dimension as compared to standard numerical method that rely on grids or meshes, allowing, for example, to find solutions to PDEs in very high dimensional settings\cite{sirignano2018dgm}.

As discussed in \cite[Sec.~8]{deluca-silverstein-torroba} an interesting alternative approach can be to replace \mbox{\textbf{Step 4}} with a direct minimization of a physical potential relevant to the problem, such as, for example, the effective potential for compactifications derived in \cite{Douglas:2009zn}. We leave a detailed analysis of this possibility to future work.

In Sec.~\ref{sub:ML3d} we will specialize and apply this algorithm to find certain classes of Einstein metrics on three-dimensional manifolds, but we stress that the method is easily generalizable to more complicated equations in higher dimensions and to include more physical fields other than the metric.  

\subsection{de Sitter compactifications of M-theory on negatively curved manifolds}\label{sub:dshyp}
In Sec.~\ref{sec:MLE}, we will study the application of the proposed ML method to the problem of finding certain negatively curved Einstein metrics on hyperbolic manifolds, known as \emph{filled geometries}. We will review these geometries extensively in Sec.~\ref{sec:MLE}, but for now we notice that this problem amounts to solving the Einstein equation
\begin{equation}\label{eq:einEq}
	R_{mn} = - (n-1) g_{mn}
\end{equation}
on non-trivial manifolds defined by appropriately gluing facets of certain polytopes. These manifolds are proven to exist in various dimensionalities. 

These features make this problem a natural testbed and benchmark for the methods discussed above. Equation \eqref{eq:einEq} is a simpler but close analogue to the full physical system \eqref{eq:Schr}, \eqref{eq:EinstInt}, and we expect that a method that is able to directly solve \eqref{eq:einEq} in high dimensions, i.e.~\emph{without relying on simplifying features of the hyperbolic manifolds}, would be a good candidate for a method to solve the full physical system in general situations. In addition, while the filled geometries are proven to exist, finding them numerically in dimension $n\geqslant 3$ is an open problem in hyperbolic geometry \cite{martelli2015hyperbolic}. 

In fact, in addition to being a natural benchmark and mathematical open problem, this geometric problem is also of direct relevance to the problem of finding de Sitter compactifications. As introduced in \cite{deluca-silverstein-torroba}, negatively curved manifolds can be used to construct dS$_4$ compactifications of M-theory, thanks in particular to the property that once supersymmetry is broken Casimir energy is automatically generated in the controllably small cycles present in the filled geometries. This large amount of (discrete) tunability allows the Casimir energy to globally compete with the classical sources (homogeneous internal $F_7$ flux) without requiring sub-Planckian cycles, and producing a meta-stable minimum of the effective potential with $\Lambda > 0$.
We thus expect that a method able to find the filled geometries in dimension seven by directly solving \eqref{eq:einEq} would be a good candidate as a method for numerically solving  the equations of motions for the dS$_4$ compactifications introduced in \cite{deluca-silverstein-torroba}. While numerical solutions are not required to extract general physical properties of these vacua, they could be useful in extracting detailed information about the internal fields, allowing to compute, for example, volumes of internal cycles.

In the next section, we discuss in detail the geometric problem associated to \eqref{eq:EinstInt} and, as a proof-of-concept of the feasibility of the ML method introduced in Sec.~\ref{sub:mlgeo}, we apply it to the three-dimensional case.

\section{Machine Learning negatively curved Einstein metrics}\label{sec:MLE}

In any dimension $n\geqslant 3$, finite volume Einstein manifolds with negative curvature can be constructed starting from an $n$-dimensional hyperbolic manifold with $k$ cusps and \emph{filling}  $k^\prime\leqslant k$ of its cusps. If $k^\prime = k$, the resulting Einstein manifold is compact. To obtain this result,  \cite{anderson-dehn, bamler2012construction} first constructed an approximate metric and then proved the existence of a suitable deformation. We start by reviewing this construction in Sec.~\ref{sub:filling}, and by describing an explicit three-dimensional hyperbolic manifold in Sec.~\ref{sub:M3}. In Sec.~\ref{sub:ML3d} we then use the ML method introduced in Sec.~\ref{sub:method} to numerically perform the deformation and obtain the final metric on these manifolds.

\subsection{Hyperbolic manifolds and Anderson-Dehn filling}\label{sub:filling}

Hyperbolic manifolds are Riemannian manifolds with constant negative sectional curvature. This strong constraint on the curvature makes this class of manifolds simpler to study with a variety of methods, allowing the construction of many examples of non-trivial manifolds in any dimension. 

An algebraic way to construct $n$-dimensional hyperbolic manifolds is to start with the hyperbolic space $\mathbb{H}^n$ and quotient it by the action of a subgroup $\Gamma$ of the isometry group of $\mathbb{H}^n$. A common choice is to start from a \emph{Coxeter polytope}, that is, a finite polytope in $\mathbb{H}^n$ whose dihedral angles divide $\pi$. These polytopes are fundamental domains of discrete reflection groups $\Gamma$.
A manifold (as opposed to an orbifold) is then obtained by identifying a suitable torsion-free subgroup $\Gamma' < \Gamma$, or, more geometrically, by explicitly identifying the facets of the polytope in an appropriate way. For more details on hyperbolic manifolds and their constructions, we refer the reader to the review \cite{martelli2015hyperbolic} and to the standard textbooks \cite{vinberg1993geometry,Ratcliffe2019FoundationsOH}.
In Sec.~\ref{sub:M3} we will review the construction of hyperbolic manifolds by using right-angled polytopes as recently developed in \cite{italiano2024hyperbolic}.

An important property of finite-volume hyperbolic manifolds is that generically they have one or more cusps,
that is, non-compact ends where the local geometry can be written as 
\begin{equation}\label{eq:cusp}
	ds^2_\text{cusp} = \frac{dr^2}{r^2} + \frac{r^2}{\rj^2} ds^2_{\mathbb{T}^{n-1}} \;,\qquad \qquad r \geqslant 0 \;.
\end{equation}
Here $r$ is a coordinate going along the cusp, $\rj$ is a real constant introduced for later convenience, and the cross-section at fixed $r$ is an $(n-1)$-dimensional torus\footnote{Other flat manifolds could arise as cusp section, but they can always be lifted to tori.}. At the infinite distance point  $r = 0$ this transverse torus shrinks to zero size. 

While hyperbolic manifolds provide a rich class of non-trivial Riemannian manifolds, the constraints imposed by the hyperbolic structure could be too restrictive for applications to gravitational physics. As an example of restriction, in dimension $n\geqslant 4$ there is only a finite number of complete hyperbolic manifolds with bounded volume \cite{wang1972topics}. More specifically, calling $\rho_n (V)$ the number of such hyperbolic manifols with volume $\leqslant V$, \cite{burger2002counting} showed that there exist two positive constants $a,b$ depending only on $n$ such that $V^{a V}\leqslant \rho_n(V) \leqslant V^{b V}$.
In addition, local regions such as \eqref{eq:cusp} can be problematic to study in low-energy approximations, since the presence of cycles shrinking to zero size can source non-perturbative effects (such as wrapping of branes) that are hard to control.

However, when solving the Einstein equations we impose weaker constraints on the curvature since only the \emph{average} extrinsic curvature, the Ricci tensor, enters in the equations, thereby allowing much larger classes of possible structures.  

Luckily, a beautiful mathematical result known as Anderson-Dehn filling bridges the world of constant extrinsic curvature to the world of constant Ricci tensor in the case of negative curvature \cite{anderson-dehn,bamler2012construction}. Specifically, in any dimension $n \geqslant 3$ \cite{thurston2022geometry,anderson-dehn,bamler2012construction}, we can cut the local geometry \eqref{eq:cusp} at $r = \rj$, choose a closed geodesic $\sigma$ in the torus cross-section and glue it to the boundary of a $D^2\times \mathbb{T}^{n-2}$, where $D^2$ is a two-dimensional disk, equipped with the euclidean AdS-BH metric
\begin{equation}\label{eq:met-BH}
    ds^2_\text{BH} = \left(\frac{dr^2}{V(r)} +V(r)d\theta^2+r^2 ds^2_{\RR^{n-2}}\right)/\Gamma \,,
\end{equation}
where \begin{equation}
    V(r) = r^2\left(1-r^{1-n}\right)\;,\qquad r \in [1,\rj]\;,
\end{equation}
and $\Gamma \cong \mathbb{Z}^{n-2}$ acts on $D^2\times \RR$ by isometries of the metric in the brackets that leave the coordinate $r$ invariant. Smoothness at the tip $r = 1$ requires $\theta$ to have periodicity $\beta \equiv \frac{4\pi}{n-1}$. Gluing the boundary of the disk to the geodesic $\sigma$ fixes $\rj$ in terms of $|\sigma|$, the proper length of $\sigma$,  via the equation 
\begin{equation}
    V(\rj)\beta^2 = |\sigma|^2\,.
\end{equation}
This procedure caps off the cusp by `killing' the closed geodesic $\sigma$ and creating a geometry that ends smoothly at  $r = 1$, the Wick-rotated horizon of the Euclidean black hole metric \eqref{eq:met-BH}. At this point, the remaining $\TT^{n-2}$ gets to a minimal size controlled by $|\sigma|$, allowing to control the minimal length of closed cycles in the obtained geometry and providing a way to avoid the problem of geometries with too small cycles. In Sec.~\ref{sub:3dDF}, we will describe this procedure in detail in a three-dimensional example. 

The piecewise metric constructed by starting from a given hyperbolic manifold $M$ and replacing the end of one of its cusp with the metric \eqref{eq:met-BH} is Einstein, but it is non-smooth at the gluing locus $\rj$.  \cite{anderson-dehn,bamler2012construction} then prove that a `nearby' smooth Einstein metric exist. 
In fact, if the starting hyperbolic manifold $M$ has $k$ cusps, this procedure can be performed independently to each of its cusps, by choosing closed geodesics $\{\sigma_i\}$, $i = 1, \dots, k$ for each of the transverse tori. If all the cusps are filled, the resulting manifold is a closed, smooth, Einstein manifold with negative curvature. Independently filling all the cusps in this way allows a great amount of tunability for the sizes of the minimal cycles in the cusps, an effect that can be exploited to tune the amount of Casimir energy building up individually in each cusp in the construction of dS$_4$ vacua reviewed in Sec.~\ref{sub:dshyp}.

Since the spectrum of closed geodesics in a given torus is unbounded above, starting from a single hyperbolic manifold with $k$ cusps, this procedure generates $\infty^k$ new Einstein manifolds. These manifolds are not isometric, and in fact have different volumes controlled by $\{|\sigma_i|\}$, accumulating at the volume of the starting hyperbolic manifold \cite{anderson-dehn}. 
An important property of both the hyperbolic manifolds and the Einstein manifolds obtained by this procedure is that they are \emph{rigid}, that is they do not admit continuous deformations \cite{besse-Einstein, anderson-dehn}. This property makes them particularly suitable for application to gravity compactifications, since it greatly simplifies the problem of moduli stabilization \cite{deluca-silverstein-torroba}.

In the next section, we describe the filling procedure in local coordinates in the three-dimensional case, constructing the approximate piece-wise Einstein metrics that we will deform in Sec.~\ref{sub:ML3d} to smooth solution of the Einstein equations. We stress that, even thought for simplicity we are starting with the three-dimensional case, the discussion above as well as the generality of the numerical method  discussed in Sec. \ref{sub:method} allow the construction of explicit negatively curved Einstein manifolds in any dimension \mbox{$n\geqslant 3$}. 

\subsubsection{The three-dimensional case in detail}\label{sub:3dDF}

We take a three-dimensional cusped hyperbolic manifold and focus on a single cusp, where the local geometry is \eqref{eq:cusp}. To fix the notation, we assume a rectangular cross-section, and we parametrize the metric on the torus at $r = \rj$ as
\begin{equation}\label{eq:origTorus}
ds^2_{\text{join}} = \dd x_1^2 + \dd x_2^2\,,\qquad x_i \sim x_i +L_i,
\end{equation}
where $L_i$ are the lengths of the two orthogonal cycles. All the quantities in this section are dimensionless: in the normalization where the Ricci scalar of a three-dimensional hyperbolic manifold is $R = -6/\ell^2$, we are working in units where $\ell = 1$.

We specialize the procedure in \cite[Sec.~3]{anderson-dehn} and \cite[Sec.~2.4]{bamler2012construction} to a specific three-dimensional example.  At $r = \rj$, we want to glue the cusp metric \eqref{eq:cusp}, with torus cross section \eqref{eq:origTorus}, to $M_\text{BH}/\Gamma \cong D^2 \times S^1$ equipped with the Euclidean-AdS black hole metric. To perform this construction, consider $M_\text{BH} \cong D^2 \times \RR$ with metric
\begin{equation}\label{eq:BH-3d}
	\tilde{ds}^2_\text{BH} = \frac{\dd r^2}{V(r)} +V(r)\dd \theta^2+r^2 \dd x_3^2 \,,\qquad\quad r \in [1, \rj],\;\quad \theta \sim \theta + \beta\;,\quad x_3 \in \RR\;.
\end{equation}
To prescribe the action of $\Gamma  \cong \ZZ$ on $M_\text{BH}$, we first describe its action on its boundary $S(\rj) \equiv M_{\text{BH}}(r = \rj) \cong S^1 \times \RR$, and we then extend this action isometrically in the interior of the disk. 

On the $\TT^2$ with metric $g_{\rj} \equiv$ \eqref{eq:origTorus}, we select a closed geodesic $\sigma$, and denote its proper length by $|\sigma|$.  We can then choose $\rj$ such that the circle size in $S(\rj)$ equals $|\sigma|$.  This requirement fixes
\begin{equation}\label{eq:fixrj}
	V(\rj) \beta^2 = |\sigma|^2
	\qquad\qquad \implies \qquad \quad \rj^2  = 1 + \frac{|\sigma|^2}{4 \pi^2} =1 + \frac{k^2 L_1^2 + L_2^2}{4 \pi^2} \;.
\end{equation}
With this identification, $S(\rj) \cong \RR^2/\langle\sigma\rangle$.  The action of the group $\Gamma \cong \ZZ $ on $S(\rj)$ consists then in translations by a vector $\vec{b}$ such that $S(\rj)/\Gamma  = \RR^{2}/\langle\sigma, b\rangle \cong (\TT^2, g_{\rj})$. To construct this action and explicitly show the isometry, consider that the original torus \eqref{eq:origTorus} can be alternatively generated by a set of generators $(\vec{\sigma}, \vec{b})$, obtained from the original ones via an SL$(2,\ZZ)$ transformation. To be concrete, we select a closed geodesic $\sigma \subset (\TT^2, g_{\rj})$ that wraps $k$ times along the $x_1$ direction when wrapping once along the $x_2$ direction, such that  $|\sigma|^2 = k^2 L_1^2+L_2^2$. In the $(x_1, x_2)$ coordinates for the universal cover $\RR^2$ of $(\TT^2, g_{\rj}$), the associated vector $\vec{\sigma} = (k L_1, L_2)$; $\vec{b}$  can then be chosen to be  $\vec{b} = (-L_1, 0)$. 

To construct the action of $\vec{b}$ on $S(\rj)$, it is useful to start from its covering \mbox{$\tilde{S}(\rj)\cong \RR^2$}. In the local coordinates $(\theta, x_3$) the metric induced from \eqref{eq:BH-3d} reads
\begin{equation}\label{eq:metSrj}
	ds^2_{\tilde S} = V(\rj) \dd\theta^2 + \rj^2 \dd x_3^2
\end{equation}
and, thanks to the identification \eqref{eq:fixrj}, $\langle\sigma\rangle$ acts by the isometry $(\theta, x_3) \mapsto (\theta, x_3) + \tilde{\sigma}$, with $\tilde{\sigma} \equiv (\beta, 0)$. 
We can then perform a coordinate change $\varphi: \RR^2 \to \RR^2$ such that $\tilde \sigma$ is mapped to $\vec{\sigma}$ and  \eqref{eq:metSrj} is mapped to \eqref{eq:origTorus}.  Explicitly, this is given by 
\begin{equation}\label{eq:map1}
	\left\{ \begin{aligned}
		x_1 &=  - \frac{L_2 \rj}{|\sigma|} x_3 + \frac{k L_1}{\beta} \theta  \\
		x_2 &=  \frac{k L_1 \rj}{|\sigma|}  x_3 + \frac{L_2}{\beta} \theta 
	\end{aligned}\right.\;.	
\end{equation}
Mapping $\vec{b}$ to the $(\theta, x_3$) coordinates using \eqref{eq:map1}, we find that it acts on \eqref{eq:metSrj} via the isometry 
\begin{equation}\label{eq:shift}
	(\theta, x_3) \mapsto (\theta, x_3) + 	\left( \frac{- k L_1 \beta}{|\sigma|^2}, \frac{L_1 L_2}{\rj |\sigma|}\right)\;.
\end{equation}
Summarizing, the action of $\Gamma$ on $S(\rj)$, with metric \eqref{eq:metSrj}, is generated by the isometry \eqref{eq:map1}.  This is also an isometry of \eqref{eq:BH-3d} when extended to act trivially on the coordinate $r$, and thus extends to an action of $\Gamma$ on $M_{\text{BH}} = D^2\times \RR$ whose quotient gives the filled cusp metric associated to $\sigma$. This action is is fixed point free and thus $M_{\text{BH}}/\Gamma$ is smooth.

Finally, we can write the metric on $M_{\text{BH}}/\Gamma$ in coordinates adapted to the original ($\TT^2, g_{\rj}$) by performing back the transformation \eqref{eq:map1} to the black hole metric \eqref{eq:BH-3d}. This results in the rather symmetric expression
\begin{equation}\label{eq:BHfin-r}
	\begin{split}
		ds^2_{\text{filled}} & = \frac{\dd r^2}{V(r)}+                       \\
		          & +\frac{1}{k^2+\alpha^2}\left[ \left(k^2 \frac{V(r)}{V(\rj)}+\alpha^2 \frac{r^2}{\rj^2}\right)\dd x_1^2+\left(\alpha^2 \frac{V(r)}{V(\rj)}+k^2 \frac{r^2}{\rj^2}\right)\dd x_2^2 +\right. \\
		          & \left.+2 k \alpha\left( \frac{V(r)}{V(\rj)}- \frac{r^2}{\rj^2}\right) \dd x_1 \dd x_2\right]\;,
	\end{split}
\end{equation}
where $\alpha \equiv L_2/L_1$ and $x_i \sim x_i + L_i$.

Summarizing, the metric \eqref{eq:BHfin-r} is glued to the original cusp metric 
\begin{eqnarray}\label{eq:cusp-fin-r}
	ds^2_\text{cusp} = \frac{\dd r^2}{r^2} + \frac{r^2}{\rj^2}\left( \dd x_1^2 + \dd x_2^2\right)\;,\qquad \qquad x_i \sim x_i + L_i \;,
\end{eqnarray} 
at a location $r = \rj$ determined in  \eqref{eq:fixrj}. 
It represents the filling obtained by smoothly shrinking a closed geodesics of length $|\sigma| =  \sqrt{k^2 L_1^2  + L_2^2}$ of the original torus, with $k$ an integer. At the end of the cusp, the remaining circle generated by \eqref{eq:map1} gets to a controllable minimal length $\ell^2(S^1)  =  \frac{L_1^2 L_2^2}{\rj^2 |\sigma|^2}$, thus removing the problem of very small cycles in the geometry by appropriate choices of $\sigma$. For $k  \gg 1$, $\rj \to \infty$, a limit in which \eqref{eq:BHfin-r} reproduces the full original cusp. 

When \eqref{eq:BHfin-r} is glued to \eqref{eq:cusp-fin-r} the resulting metric is continuous and Einstein, but it is not smooth at $ r = \rj$. This metric will be the starting point of our numerical analysis in the next section, where we will deform it to a smooth solution of the Einstein equation. While in three dimensions 	an Einstein metric is locally hyperbolic, this is not the case in higher dimensions, and this procedure produces a genuine piecewise Einstein metric.

\subsection{An explicit 3d manifold}\label{sub:M3}

As we briefly introduced in Sec.~\ref{sub:filling}, one way to construct $n$-dimensional hyperbolic manifolds is to start from polytopes in $\mathbb{H}^n$ and prescribe an appropriate identification among their facets. In the recent work \cite{italiano2024hyperbolic},  the authors introduce a general method to perform this procedure in $3\leqslant n \leqslant 8$ by employing a special class of \emph{right-angled} hyperbolic polytopes known as $P_n$ \cite{potyagailo2005right, everitt2012right}, thereby constructing explicit examples of finite-volume cusped hyperbolic manifolds $M_n$ in dimensions up to 8. While six and seven are the more natural dimensionalities to consider for four-dimensional compactifications of string/M-theory, their construction method is general, and in this section we review the simpler three-dimensional case in detail, expanding upon the description in\cite[App.~A.1]{deluca-silverstein-torroba}.

\subsubsection{The UHS realization}\label{sub:uhs}

For definiteness, we will construct the polytopes in the Upper-Half Space (UHS) model of the hyperbolic space, described in three dimension by the metric 
\begin{equation}\label{eq:UHS}
	ds^2_{\text{UHS}} = \frac{1}{z^2}(\dd z^2 + \dd x_1^2 + \dd x_2^2)\;,\qquad x_i \in \RR\;,\; z \in \RR^+\;.
\end{equation}
Since this model is conformally equivalent to flat space, right-angled hyperbolic polytopes are right-angled also as euclidean polytopes. To construct them we can only use totally geodesics facets, which in this model are either vertical planes or hemispheres with centers on the $z = 0$ plane.

\begin{figure}[!ht]
	\centering
	\includegraphics[width=5cm]{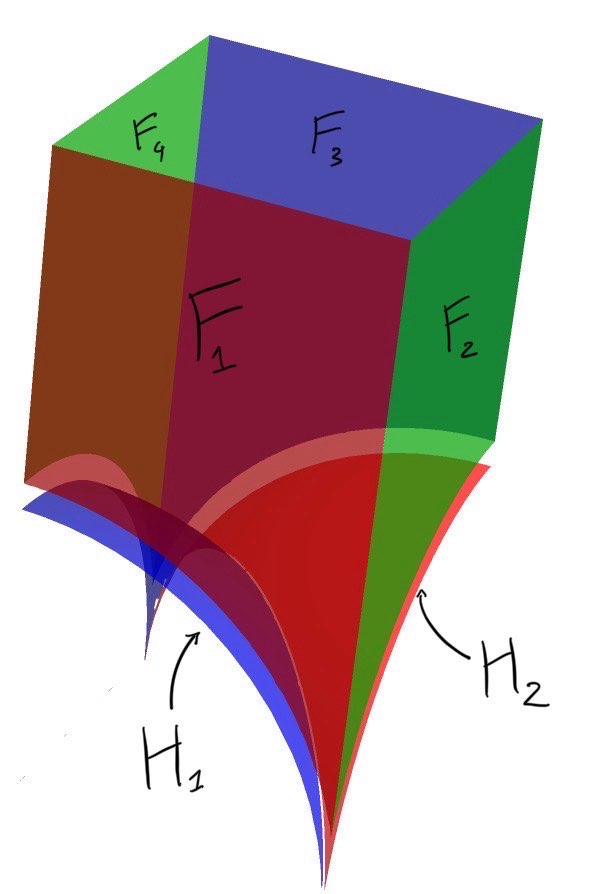}
	\caption{\small An embedding of $P_3$ in the UHS model \eqref{eq:UHS} with $z$ the vertical coordinate. This polytope is composed of 6 totally geodesics facets: four vertical walls and two hemispheres centered at $z = 0$. In this picture the coordinate $z$ has been cut-off at a finite volume, but the polytope extends up to $z = \infty$, while remaining of finite volume with respect to the metric \eqref{eq:UHS}.}\label{fig:P3_colored} 
\end{figure}

The building block for the construction of the hyperbolic manifold $M_3$ using the general prescription in \cite{italiano2024hyperbolic} is the right-angled polytope known as $P_3$. This polytope has 6 facets, and in the UHS model it can be constructed by taking four vertical planes
\begin{equation}
	\quad \left\{ \begin{aligned}
		F_1 & : & x_1 = 0, x_2 \in (0,L), z \in (0,\infty) \\
		F_2 & : & x_1 \in(0,L), x_2 = L, z \in (0,\infty)  \\
		F_3 & : & x_1 = L, x_2 \in (0,L), z \in (0,\infty) \\
		F_4 & : & x_1 \in(0,L), x_2 = 0, z \in (0,\infty)  \\
	\end{aligned}\right.
\end{equation}
and two hemispheres $H_1$ and $H_2$ with radius equal to $L$ and with centers $C_1$ and $C_2$ respectively at
\begin{equation}
	C_1 : (0,0,L)\;,\qquad \qquad C_2: (0,L,0)\;.
\end{equation}
Here $L$ is an overall length scale that we will henceforth set to 1. 
These facets intersect with right angles, and the polytope $P_3$ is the interior region, together with its boundary. Fig.~\ref{fig:P3_colored} shows a representation of $P_3$.

	$P_3$ can be identified with the quotient $\mathbb{H}^3/\Gamma$ with $\Gamma$ being the Coxeter group generated by the reflections along the facets of $P_3$. To construct a manifold, we need to identify a subgroup 
$\Gamma^\prime \leq \Gamma$ whose action on $\mathbb{H}^3$ is free. A standard method consists in prescribing a \emph{coloring} of the facets of the right-angled polytope and a corresponding reflection group. 
More formally \cite[Sec.~1.1]{italiano2024hyperbolic}, a \emph{c-coloring} of a polytope $P$ is the assignment of a color (taken from some fixed set of $c$ elements) to each facet of $P$, such that incident facets have distinct colors. A coloring on $P$ induces a homomorphism $\Gamma \to \ZZ_2^c$, by denoting with a one-hot encoded bit string whether the corresponding face is colored with a given color. It can be shown that the kernel $\Gamma^\prime \triangleleft \Gamma$ acts freely on $\mathbb{H}^n$ thus defining a manifold $M = \mathbb{H}^n/\Gamma^\prime$ composed by $2^c$ copies of $P$. More geometrically, for each color $c$, choose a single facet $F_c$ colored with that color. For each bit string $\xi \in \ZZ_2^c$ denote by $P_{\xi}$ the copy of $P$ obtained by reflecting $P$ once along the corresponding facet $F_c$. We then identify each facet $F_i$ of $P_\xi$ with the same facet of $P_{\xi+\xi_i}$ where $\xi_i$ is the bit-string encoding the color of $F_i$. Notice that while the trivial coloring, assigning a different color to each facet of $P_n$, is always available, the volume and the combinatorial complexity of the corresponding manifold grows exponentially with the dimension, and can thus quickly become intractable in high dimensions. 

The smallest coloring of $P_3$ consists of $3$ colors, and all the three-colorings are isomorphic. In our notation above for the facets, we prescribe this coloring by using colors  (Red, Green, Blue) and coloring  the facets as
\begin{center}
	\begin{tabular}{|c|c|c|c|c|c|}
		\hline
		$F_1$              & $F_2$                & $F_3$               & $F_4$                & $H_1$               & $H_2$              \\
		\hline
		\cellcolor{red!25}$\mathcal{R}$& \cellcolor{green!25}$\mathcal{G}$&\cellcolor{blue!25}$\mathcal{B}$& \cellcolor{green!25}$\mathcal{G}$&\cellcolor{blue!25}$\mathcal{B}$& \cellcolor{red!25}$\mathcal{R}$\\
		\hline
	\end{tabular}\;,
\end{center}
as depicted in Fig.~\ref{fig:P3_colored}.

\begin{figure}[!ht]
	\centering
	\includegraphics[width=13cm]{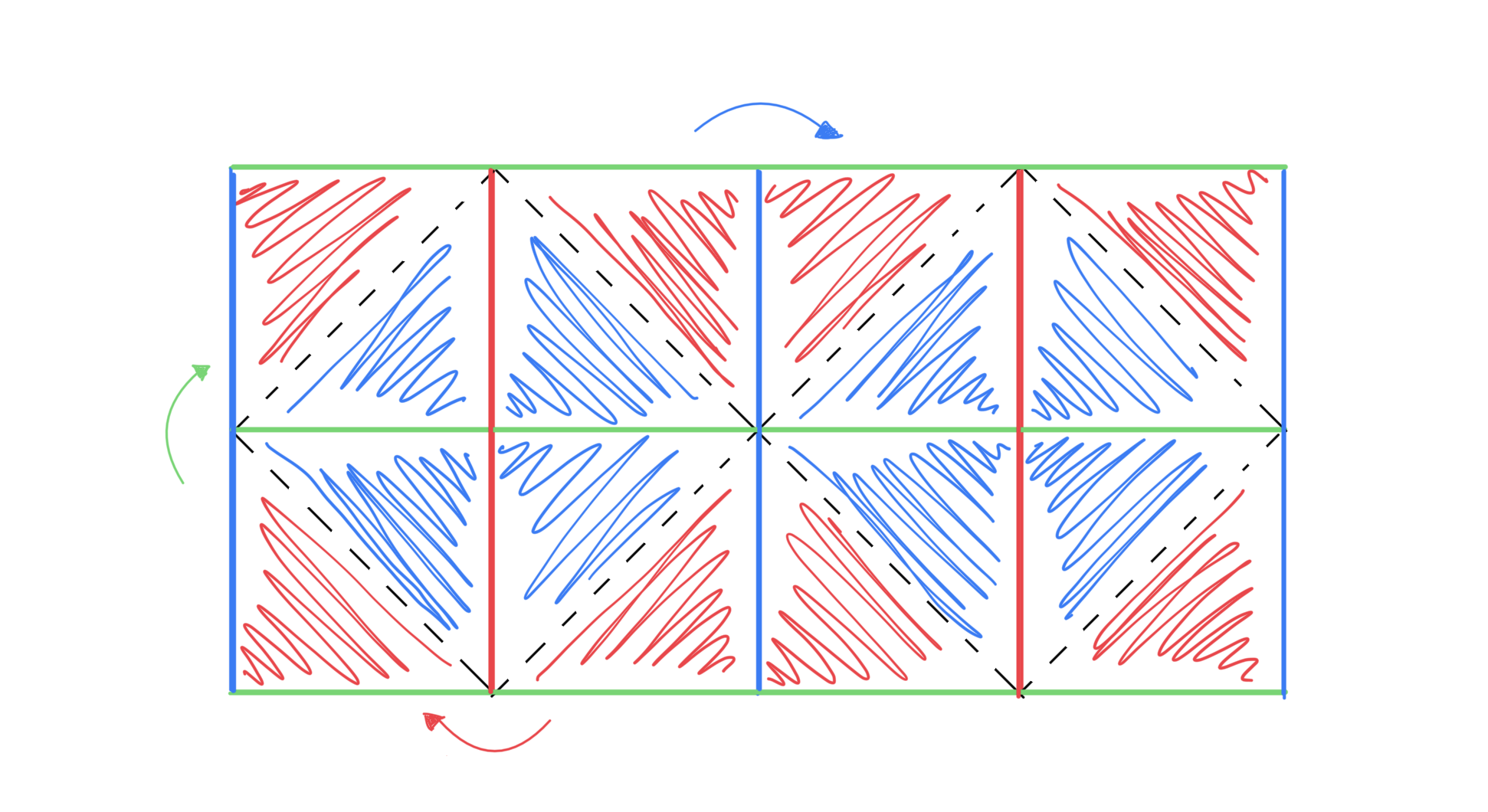}
	\caption{A view from the top of the larger polytope $P_3^8$ obtained by reflecting according to the reflections \eqref{eq:refl}}\label{fig:P38} 
\end{figure}

To generate $P_3^8$ we have to reflect once with respect to each color, and we choose the facets $F_1$, $F_2$, and $F_3$, in this order. Explicitly, the corresponding reflections are
\begin{equation}
	\quad \left\{ \begin{aligned}
		R_1 & : & x_1 & \to -x_1   \\
		R_2 & : & x_2 & \to 2L-x_2 \\
		R_3 & : & x_1 & \to 2L-x_1 \\
	\end{aligned}\right.\label{eq:refl}
\end{equation}
Figure \ref{fig:P38} shows the obtained $P_3^8$.

To implement the numerical method it suffices to  randomly draw points from the bulk of $P_3^8$ and from its boundaries. To do so, we generate a point on $P_3$ and act with the above reflections, keeping all the intermediately generated points.
Similarly, we choose $F_1$, $F_2$, $F_3$ as the facets $F_c$ representing each of the colors, and we identify the 28 external facets of $P_3^8$ according to the method discussed above. We provide more details about the 14 identified facets and an algorithm to draw points from them in App.~\ref{app:detM3}. 

Upon appropriately identifying the external facets of $P_3^8$, we finally end up with the finite volume manifold $M_3$. This manifold has 3 cusps, two at $z = 0$ and one at $z = \infty$. In the next section, we will deform the hyperbolic metric of the UHS to cap off the cusp at $z = \infty$. A similar procedure can be applied to all the cusps, for example by first performing an hyperbolic isometry that maps any given cusp to $z = \infty$.

\subsubsection{The piece-wise metric}\label{sub:piecewise}
Due to the identifications induced by the coloring, for $z\geqslant 1$ the hyperbolic metric on $M_3$ reads
\begin{equation}\label{eq:UHS-zg1}
	ds^2 = \frac{1}{z^2}(\dd z^2 + \dd x_1^2 + \dd x_2^2)\;,\qquad x_1 \in [-1, 3]\;,\; x_2 \in [0,2]\;,\;z \geqslant 1. 
\end{equation}
With the coordinate change $z = \frac{\rj}{r}$,
this agrees with the local cusp metric \eqref{eq:cusp}, with $r = \rj$ mapped to $z = 1$. At this locus, the torus metric has the form \eqref{eq:origTorus}, with $(L_1,L_2) = (4,2)$. Following the analysis in Sec.~\eqref{sub:3dDF} we can thus choose a closed geodesic $\sigma$ that wraps $k$ times along the $x_1$ direction while wrapping once along $x_2$,  and obtain the corresponding filled metric \eqref{eq:BHfin-r}.
Mapping it back to the coordinate $z$, and defining 
\begin{equation}\label{eq:Wz}
	W(z) \equiv V(r= z^{-1}\rj ) = z^{-2}\rj^2 -1\;,
\end{equation}
we finally obtain on $M_3$ the piece-wise metric
\begin{equation}\label{eq:piecewise}
	\overline{ds}^2_{M_3} =  \left\{\begin{array}{lccr}
		z^{-2}(\dd z^2 + \dd x_1^2 + \dd x_2^2) &\quad&\hfill&0\leqslant z \leqslant 1\\ \\
		\begin{aligned}
		  &\rj^2\frac{dz^2}{z^4W(z)}+         \frac{1}{k^2+\alpha^2}\left[ \left(k^2 \frac{W(z)}{W(1)}+\frac{\alpha^2}{z^2}\right) \dd x_1^2+ \right.                                                                     \\
		          &\left. +\left(\alpha^2 \frac{W(z)}{W(1)}+\frac{k^2}{z^2} \right)\dd x_2^2
		           +2 k \alpha\left( \frac{W(z)}{W(1)}- \frac{1}{z^2}\right) \dd x_1 \dd x_2\right]
	\end{aligned} &\quad& \hfill& 1\leqslant z \leqslant \rj
\end{array}\right.
\end{equation}
where $\rj$ is determined from \eqref{eq:fixrj} and $\alpha = L_2/L_1$.
\eqref{eq:piecewise} defines a metric on $M_3$, once the cooordinates $x_i$ are identified according to the discussion in the previous section. This metric is continuous and everywhere Einstein (and thus locally hyperbolic in 3 dimensions), but it is not smooth at $ z = 1$. \cite{anderson-dehn,bamler2012construction} perform a two-step analysis where they first manually deform  it into a smooth (but non-Einstein) metric in a region around the gluing point, and then prove that a `nearby' smooth and Einstein metric exist. 

In the numerical procedure in Sec.~\ref{sub:ML3d} we will implictly define a smooth approximation of \eqref{eq:piecewise} by pre-training a set of Neural Networks to match \eqref{eq:piecewise}, without yet enforcing the Einstein condition. Since with appropriate choices for the architecture Neural Networks are smooth functions, this procedure will fail to match \eqref{eq:piecewise} exactly, thereby defining a smooth approximation to an Einstein metric on $M_3$, which we will then deform into a smooth solution of the Einstein equation. 

\subsubsection{The boundary conditions}\label{sub:bry}

To construct $M_3$ we pairwise identify the external facets of $P_3^8$. When the metric on $P_3^8$ is the hyperbolic metric,  this is guaranteed to produce a smooth metric on $M_3$ by the arguments reviewed in Sec.~\ref{sub:uhs}. When deforming away from the hyperbolic metric we need to impose that the deformed metric is still continuous and differentiable at the interface where the facets are identified. This is tantamount to requiring that the intrinsic geometry of the interface, as induced by the ambient metric, is well-defined. Explicitly, this requires the intrinsic metric and the extrinsic curvature of the interface to agree on its two sides, as we will briefly review. We refer the reader to the standard textbooks \cite[Chap.~3]{poisson2004relativist}, \cite[App.~D]{carroll2019spacetime} for more extensive discussion of  these conditions.

Denote by $(A_i, B_i)$ the pairs of external facets of  $P_3^8$ that have to be identified, and with $\Sigma_i$ the corresponding hypersurface in $M_3$, with intrinsic coordinates $\sigma$. 
We have two maps $\alpha_i$ and $\beta_i$ from $\Sigma_i$ to $M_3$, which embed $\Sigma_i$ as $A_i$ and $B_i$, respectively.
To impose that the intrinsic metric on $\Sigma_i$ is well defined,  we require that it agrees on the two sides. In local coordinates:
\begin{equation}\label{eq:hab}
	h_{ab}(\sigma)\rvert_{A_i} = h_{ab}(\sigma)\rvert_{B_i} \hfill\implies\hfill 
	g_{m n}(x) (e_{A_i})^m_a(\sigma)(e_{A_i})^n_b(\sigma) = g_{m n}(\hat{x}) (e_{B_i})^m_a(\sigma)(e_{B_i})^n_b(\sigma)\;,
\end{equation}
where $a,b$ are indices on $\Sigma_i$, $x = \alpha_i(\sigma)$, $\hat{x} = \beta_i(\sigma)$, and
\begin{equation}
	(e_{A_i})^m_a(\sigma) \equiv \frac{\partial x^{m}}{\partial \sigma^a}\;, \qquad (e_{B_i})^m_a(\sigma) \equiv \frac{\partial \hat{x}^{m}}{\partial \sigma^a}\;.
\end{equation}
Notice that when the coloring is chosen such that the facets are only mapped by reflections with respect to vertical planes, $\hat x$ and $x$ are connected by a simple affine transformation as $\hat{x}_i = J_i x_i + \Delta_i$ so that $(e_{B_i})^\mu_a(\sigma) = (J_i)^\mu_\nu(e_{A_i})^\nu_a(\sigma)$. In App.~\ref{app:detM3} we collect the maps and the Jacobians explicitly for the facets of $P_3^8$ with the coloring described in Sec.~\ref{sub:M3}. 

Similarly, smoothness is imposed by requiring that the extrinsic curvature is continuos: as in the Israel junction conditions, a jump in extrinsic curvature would not be compatible with the Einstein equations unless sourced by a localized stress-energy tensor. 

Applying the same analysis, we get in local coordinates the condition
\begin{equation}\label{eq:Kab}
	K_{ab}(\sigma)\rvert_{A_i} = \hat{K}_{ab}(\sigma)\rvert_{B_i} \hfill\implies\hfill  K_{m n}(x) (e_{A_i})^m_a(\sigma)(e_{A_i})^n_b(\sigma) = \hat{K}_{m n}(\hat{x}) (e_{B_i})^m_a(\sigma)(e_{B_i})^n_b(\sigma)\;,
\end{equation}
where $K_{mn}(x) \equiv \nabla_{(m} \eta_{n)}(x(\sigma))$ with $\eta^m$ the unit normal vector to $A_i$, with respect to the metric $g_{mn}(x)$, all evaluated at $x = \alpha_i(\sigma)$. Similarly, $\hat{K}_{mn}(\hat x) \equiv \nabla_{(m} \hat{\eta}_{n)}(\hat x(\sigma))$, where $\hat \eta^m$ is the unit normal to $B_i$, with respect to the metric $g_{mn}(x)$, evaluated at $x = \hat{x} = \beta_i(\sigma)$.
To impose this condition correctly, the direction of the normals has to be chosen consistently for each pair of identified facets. A simple way to algorithmically fix the sign of $\hat \eta$ with respect to the sign of $\eta$ is to map back $\hat{\eta}$ from $B_i$ to $A_i$ as  $\tilde{\eta}_\mu (x)\equiv J^\nu_\mu \hat{\eta}_\nu (\hat x)$ and impose $\tilde \eta \cdot \eta = 1$.
Notice that in the hyperbolic metric \eqref{eq:UHS}, the condition \eqref{eq:Kab} is trivially satisfied since both the vertical planes and the hemispheres have zero extrinsic curvature. 

\subsection{The Machine Learning method}\label{sub:ML3d}

We are now ready to specialize the Machine Learning method described in section \ref{sub:method} to the problem of finding Einstein metrics on the three-dimensional Einstein manifolds introduced in Sec.~\ref{sub:M3}, starting from the piece-wise configuration described in Sec.~\ref{sub:piecewise}. 
\subsubsection{The algorithm}

We follow the procedure in Sec.~\ref{sub:method} with the goal of finding Einstein metrics on $M_3$ where the cusp at $z = \infty$ has been filled. Specifically, we solve the equations
\begin{equation}\label{eq:Rmn}
	\EE: R_{mn} + \frac{2}{\delta} g_{mn}  = 0\;,
\end{equation}
where we allowed the presence of an arbitrary constant conformal factor $\delta$. (With this rescaling the purely hyperbolic solution in the UHS is $ds^2 = \delta\times\eqref{eq:UHS}$.)

We will solve equation \eqref{eq:Rmn} on $\Omega  = P_3^8$. The boundary conditions are the continuity equation \eqref{eq:hab} and the matching of the extrinsic curvature \eqref{eq:Kab}:
\begin{equation}\label{eq:Bexpl}
	\left\{\begin{array}{ll}
		\BB_{1, a}: &g_{m n}(x) (e_{A_i})^m_a(x)(e_{A_i})^n_b(x) = g_{m n}(\hat{x}) (e_{B_i})^m_a(\hat{x})(e_{B_i})^n_b(\hat{x})\\
		\BB_{2,a}:  &K_{m n}(x) (e_{A_i})^m_a(x)(e_{A_i})^n_b(x) = \hat{K}_{m n}(\hat{x}) (e_{B_i})^m_a(\hat{x})(e_{B_i})^n_b(\hat{x})
	\end{array}\right.
\end{equation}
where $a = 1, \dots, 14$ runs over all the pairs of identified boundary components and $x$ and $\hat x$ are the identified points in a given pair. 

For this simple three-dimensional problem, we did not find necessary to fix the gauge constraint $\xi = 0$, so we do not have extra constraints $\mathsf{C}$ to impose. 

We can now follow the method \ref{sub:method} step-by-step. 

\begin{enumerate}
	\item[] \textbf{Step 1}. We parametrize the metric as 
	\begin{equation}\label{eq:metNN}
		g_{mn}(z,x_1,x_2; \{\theta^{(i)}\}) \equiv \left(
		\begin{array}{ccc}
		\frac{\mathcal{N}_0(z,x_1, x_2 ; \theta^{(0)})}{z^2 (\rj-z)} & \mathcal{N}_4(z,x_1, x_2 ; \theta^{(4)}) & \mathcal{N}_5(z,x_1, x_2 ; \theta^{(5)}) \\
		\mathcal{N}_4(z,x_1, x_2 ; \theta^{(4)}) & \frac{(\rj-z)\mathcal{N}_1(z,x_1, x_2 ; \theta^{(1)})}{z^2} & \mathcal{N}_3(z,x_1, x_2 ; \theta^{(3)}) \\
		\mathcal{N}_5(z,x_1, x_2 ; \theta^{(5)}) & \mathcal{N}_3(z,x_1, x_2 ; \theta^{(3)}) & \frac{\mathcal{N}_2(z,x_1, x_2 ; \theta^{(2)})}{z^2} 
		\end{array}
		\right)\;,
		\end{equation}
		where $\mathcal{N}_i : P_3^8 \to \RR$ are 6 fully connected neural networks. More details about the architecture can be found in App.~\ref{app:detML}. The parametrization in  \eqref{eq:metNN} is chosen such that the neural networks do not have to learn the coordinate singularities at $z \to 0$ and $z \to \rj$, but only the smooth parts. 
		\item[] \textbf{Step 2}. We randomly sample points from $P_3^8$ and its boundary using as probability density the volume form of the hyperbolic metric, in order to focus the training on reducing the errors in regions that contribute mostly to the volume. In addition, to reduce more quickly the error in problematic regions, we identify a fixed fraction of the sampled points that had the highest error at the previous iteration, as measured by the training loss. This subset of points is kept fixed and not resampled for a certain specified number of iterations. This sampling scheme is simple to implement, does not require extra function evaluations, and in our experiments appeared to be effective at accelerating convergence since it dynamically weighs more regions with higher errors. 
		\item[] \textbf{Step 3} (pre-training). We initialize the parameters $\Theta \equiv \{\theta^{(i)}\}$ such that $\mathcal{N}_4 = \mathcal{N}_5 = 0 $, and we pre-train $\mathcal{N}_{0,\dots, 3}$, starting from a random initialization, in a way such that \eqref{eq:metNN} would reproduce $\delta\times \eqref{eq:piecewise}$. Specifically, we choose a loss function that imposes both the matching of the metrics, as well as their first and second derivatives in $z$. Schematically, we use
		\begin{equation}
			\begin{split}
				\mathcal{L}_{\text{pre}}(\Theta) &\equiv \sum_{x\sim \rho} \left| g_{mn}(z,x_1, x_2, \Theta) - \delta \bar{g}_{mn}(z,x_1, x_2)\right| + \\ 
					&+\sum_{x\sim \rho} \left| \partial_zg_{mn}(z,x_1, x_2, \Theta) - \delta \partial_z\bar{g}_{mn}(z,x_1, x_2)\right|\\
					&+ \sum_{x\sim \rho}\left| \partial_z^2g_{mn}(z,x_1, x_2, \Theta) - \delta \partial_z^2\bar{g}_{mn}(z,x_1, x_2)\right|\\
			\end{split}
		\end{equation} 
		where the norm is an element-wise smooth $L^1$ norm. We notice that this problem cannot be solved exactly due the neural network being differentiable functions, while $\eqref{eq:piecewise}$ is not differentiable at $z = 1$. We thus stop the training after a certain threshold for the error is achieved. At the end of this procedure, $g_{mn}(z,x_1, x_2, \Theta)$ represents a smooth approximation to the filled geometry on $M_3$.
		\item[]  \textbf{Step 4} (training). Starting from the pre-trained network, we then continue the training using the loss \eqref{eq:loss} with $\EE$ as in \eqref{eq:Rmn}, $\BB$ as in \eqref{eq:Bexpl} and $\mathsf{C} = 0$. We use as norm the absolute, component-wise, relative error,  and the same the sampling strategy discussed above. 
\end{enumerate}

We present the results in the next section.

\subsubsection{Results}\label{sec:res}
\begin{figure}[!ht]
    \centering
    \begin{minipage}[t]{0.5\textwidth}
        \centering
        \includegraphics[width=1.0\linewidth]{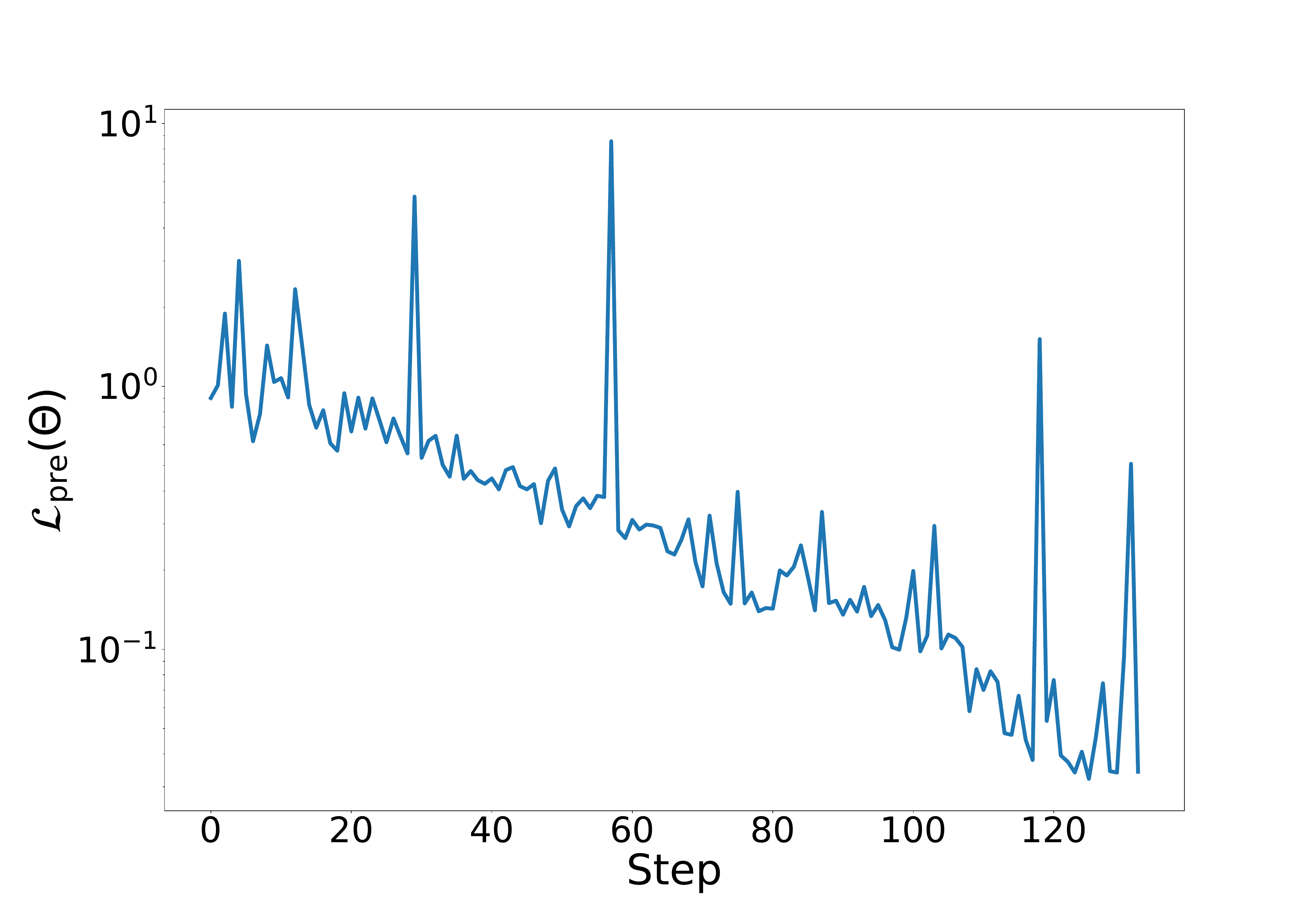}
    \end{minipage}%
    \begin{minipage}[t]{0.5\textwidth}
        \centering
        \includegraphics[width=1.0\linewidth]{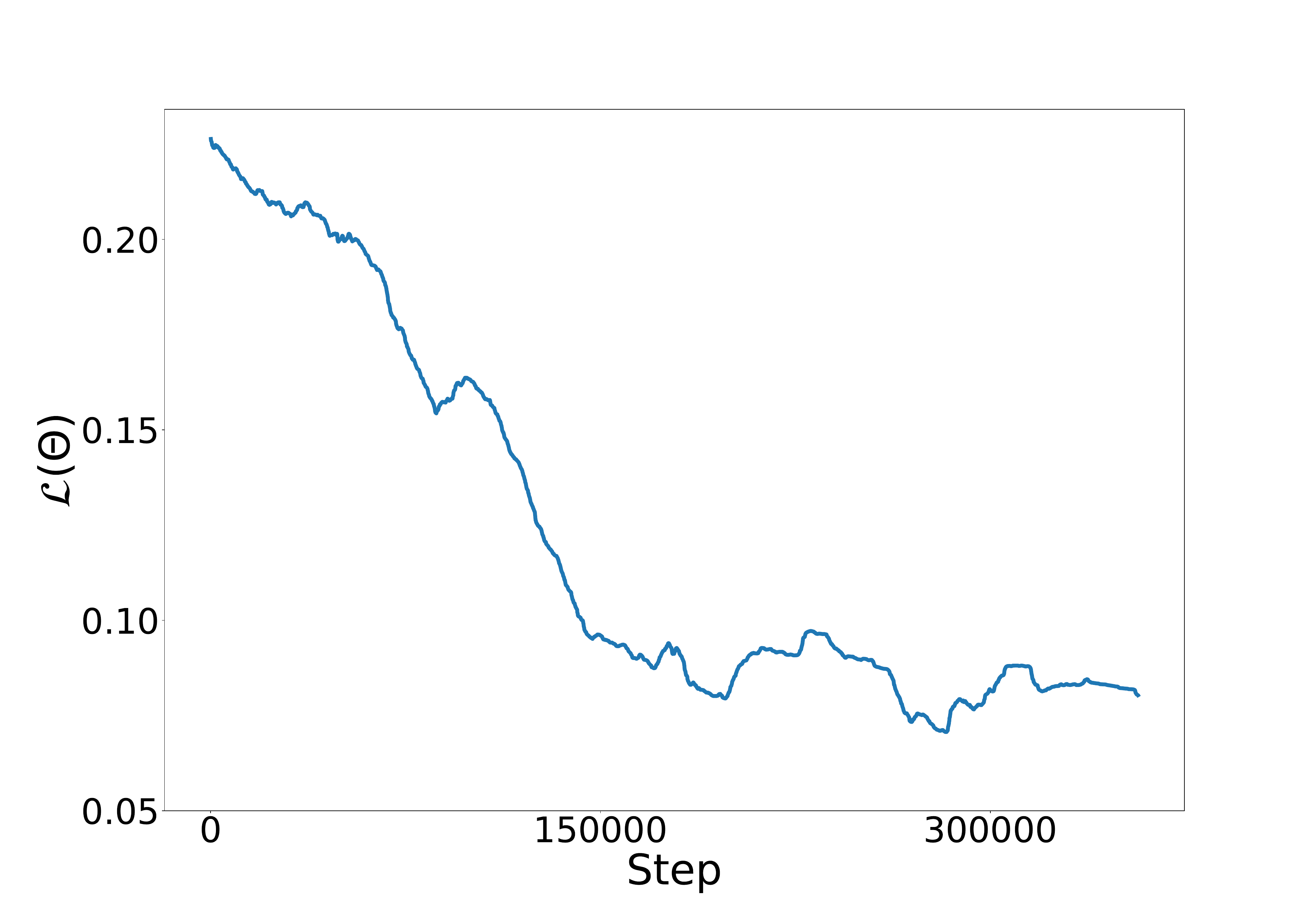}
    \end{minipage}
    \caption{\small A typical training run. Left: pre-training loss when training to approximate the piece-wise solution starting from a random initialization, right: training loss (running average) when training on the full loss starting from the pre-trained network. The pre-training is early-stopped when the pre-training loss reaches a certain threshold, as its role is only to serve as a good initialization for the actual training run.}
    \label{fig:training_run}
\end{figure}

Starting from a random initialization we pre-train until the pre-training loss gets to fixed threshold and then continue training from there by switching to a loss that encodes the full set of equations of motions and boundary conditions. A typical training run is illustrated in Fig.~\ref{fig:training_run}. 

To understand the quality of a training run, we need to define appropriate error measures. Lacking a ground truth, defining precise pointwise error measures is not immediate, and we adopted the following measure. The equations we are solving are of the form $T^1_{\mu \nu}(x) = T^2_{\mu \nu}(x)$ where $T^{(1)}$ and $T^{(2)}$ are tensors, $\mu \nu$ are appropriate indices and $x$ denotes points on the bulk and/or on the boundaries of $P_3^8$. Specifically, for the Einstein equation \eqref{eq:Rmn}, $T^{(1)} = R_{mn}$ and $T^{(2)} = 2\delta^{-1} g_{mn}$, with $m,n \in [1,2,3]$, while for the boundary conditions the tensors are the two sides of \eqref{eq:hab} and \eqref{eq:Kab}, with $a,b \in [1,2]$. We define the Local Relative Error as
\begin{equation}
\text{LRE}[T_1, T_2](x)\equiv  \frac{1}{n^2} \sum_{\mu,\nu = 1}^{n} \frac{|T^{(1)}_{\mu \nu}(x)-T^{(2)}_{\mu \nu}(x)|}{|T^{(1)}_{\mu \nu}(x)|+|T^{(2)}_{\mu \nu}(x)| + \epsilon}\;,
\end{equation}
where $n $ denotes the range of the indices and $\epsilon = 10^{-12}$ is a regularization constant. We want this error to be as small as possible, but in particular, we want the error to be smaller in regions where the volume density is larger. We thus define its average (ARE) as
\begin{equation}
	\text{ARE}[T_1, T_2] \equiv \frac{\int \sqrt{g(x)}  \text{LRE}[T_1, T_2](x)\,\dd x}{\int\sqrt{g(x)} \dd x}\;.
\end{equation}
These integrals are defined either on the bulk or on the boundary depending on the considered tensor, with $\sqrt{g(x)}$ the volume form evaluated at those points.  To compute the volume form we could use either the fiducial starting piece-wise metric or the computed metric. Given their similarity, to avoid degenerate situations we opted to use the fixed piece-wise metric.\footnote{While we have not observed this, it could happen otherwise that the Neural Networks try to minimize this error by concentrating all the volume in a small region with small error, while the error remains large elsewhere.}

To construct a baseline for our method, following \cite{anderson-dehn,bamler2012construction}, we manually define a smooth deformation of the piece-wise metric \eqref{eq:piecewise} by replacing $W(z)$ with an appropriate smooth interpolating function. The resulting metric is smooth but not Einstein in a localized region around the gluing locus. Different interpolations have different trade-offs between errors in the boundary conditions versus errors in the Einstein equation. We collect more details about the interpolations in App.~\ref{app:detML}. 

\begin{table}[ht!]
	\centering
	\begin{tabular}{l|ccc|c}
	\toprule
	\textbf{Metric} & \textbf{Einstein eq.} & \textbf{continuity} &  \textbf{ extr.~curv.} & \textbf{Avg. ARE}\\ 
	\midrule
	Interpolation 1        &        0.$145$        & $\sim 0$     &  $0.167$        & $0.104$  \\ 
	Interpolation 2        &        $0.121$        & $0.145$     &  $0.511$   & $0.259$       \\ 
	
	Trained        &  ${(9.7 \pm 0.5) \times 10^{-3}}$                 & ${(5\pm 3)\times 10^{-5}}$     &  ${(2\pm 1) \times 10^{-4}}$  & $(3.3 \pm 0.1)\times 10^{-3}$      \\ 
	\bottomrule
	\end{tabular}
	\caption{\small Average Relative Errors on the Einstein equations and the two components of the boundary conditions, together with their average, for a filling with $k= 5$. The manual interpolations, defined in \eqref{eq:interpolations}, have different trade-offs for the errors in the boundary conditions and in the Einstein equations (cf.~Fig.~\ref{fig:ricci-scalar-comparison}). The confidence intervals quoted here refer to the fact that for the trained network the ARE is computed via a Monte Carlo integration.}
	\label{tab:AREs}
	\end{table}
Table \ref{tab:AREs} compares the different components of the errors of a trained network versus the baselines. The `trained' row in the Table refers to a run with the optimizer $\mathsf{ECDSep}$ introduced in \cite{ECDSep}, with the rescaling of the hyperparameters proposed in \cite{wip-LHC}. We performed a limited hyperparameter scan both for this optimizer and comparing with more standard optimizers like Adam \cite{kingma2014adam}. For simplicity, we chose the balancing coefficients in \eqref{eq:loss} as $\gamma_{\mathsf{E}} = \gamma_{\mathsf{B}} = 1$ and, consistently with this choice, we found that different successful runs, both within an optimizer and across them, display different trade-offs among the AREs in the boundary conditions and the ARE in the Einstein equation. However, the average ARE turned out to be $\mathcal{O}(10^{-3})$ for the best runs of each optimizer, suggesting that this three-dimensional problem is not hard enough to distinguish among their performances. The result reported in the table is for one of the $\mathsf{ECDSep}$ runs that showed smaller error on the boundary conditions than in the Einstein equation.  More details about the training and hyperparameter tuning are collected in App.~\ref{app:ML}.

From Table \ref{tab:AREs} we can see that the learned metric, which depends on all the directions, is able to reduce the error simultaneously on the Einstein equations and all the boundary conditions by many orders of magnitude, suggesting convergence to the true solution. In this work and in the associated code we did not target high accuracy, and it would be interesting to see how small these errors can be made with more accurate techniques.

\begin{figure}[!ht]
    \centering
        \includegraphics[width=1.0\linewidth]{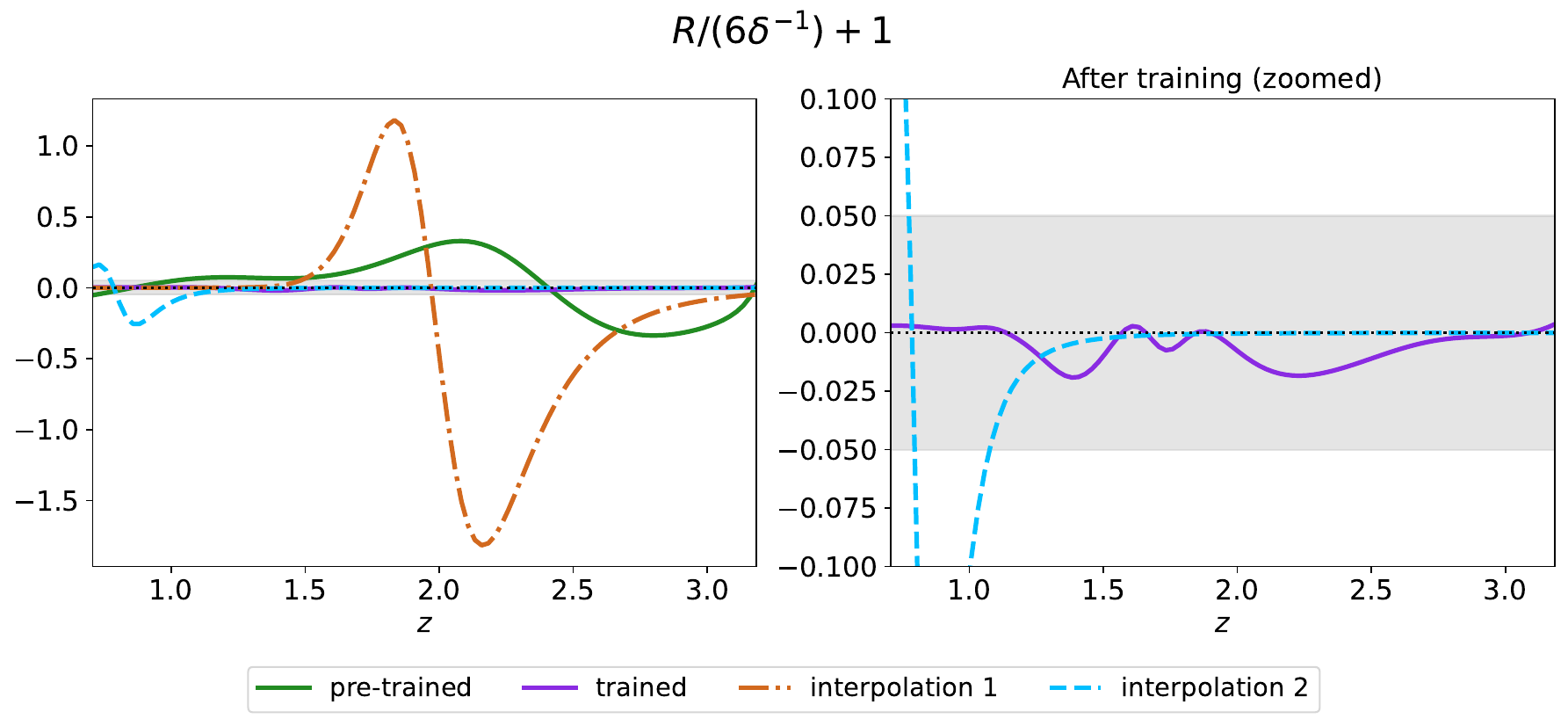}
    \caption{\small Percentage deviation of the Ricci scalar from its true value for a $k = 5$ filling, evaluated on the slice $x_1 = x_2 = \frac12$, after the pre-training phase (green) and after the training phase of a typical training run (purple), compared to two different manual smooth interpolations of the piece-wise Einstein metric. Shaded area shows 5\% error threshold.  }
    \label{fig:ricci-scalar-comparison}
\end{figure}

To guide a visual understanding of the errors, we plot in Fig.~\ref{fig:ricci-scalar-comparison} the percentage deviation of the Ricci \emph{scalar} from its true value, on a slice at fixed $(x_1, x_2)$, for a typical training run. In App.~\ref{app:detML} we compare all the components of the Ricci tensor.

\section{Conclusions and outlook}\label{sec:concl}

In this paper we initiated a study of the viability of Machine Learning methods as tools for solving the equations of motion for general warped gravity compactifications. 
We demonstrated their applicability on the proof-of-concept problem of finding certain Einstein metrics on three-dimensional Riemannian manifolds.  In Sec.~\ref{sec:res} we started from an approximate solution, which is often available and constructable with standard pertubative or patchwise numerical methods, and showed that the algorithm proposed in Sec.~\ref{sub:mlgeo} is able to converge very close to the exact solution, reducing the initial error by many orders of magnitude.

In this work, we did not target high accuracy nor we tried optimize the code for high performance. It would be interesting to explore how fast and accurate these methods can be made, using techniques such as just-in-time compilation (as employed e.g.~in the Calabi-Yau case in \cite{Gerdes:2022nzr}), higher numerical precision, and switching to higher order methods at the end of training in order to converge to the bottom of the basin more precisely. The importance of these techniques depends on the scale and on the accuracy required for the specific physics application.

An immediate generalization of the work presented in this paper would be to apply our methods and code to a higher-dimensional geometric setup, where it would lead to the explicit construction of novel Einstein metrics. We started this study in \cite{4d} and we hope to report on this soon. 

Ultimately, the goal of this approach is to develop a general method for solving the full system of semi-classical equations of motion for general warped compactifications in six or seven dimensions, reviewed in Sec.~\ref{sec:eoms}. A first natural target for this program is the detailed numerical construction of the dS$_4$ compactifictions reviewed in Sec.~\ref{sub:dshyp}. In particular, these compactifications can be constructed on higher-dimensional generalizations of the manifolds studied in this work, using as a starting point the approximate solutions derived in \cite{deluca-silverstein-torroba} and constructed via an approximation scheme that deals with the inhomogeneities of the internal fields by decomposing the internal manifold into different patches where different effects dominate. These approximate solutions can be used as a starting point for the Machine Learning method, analogously to how we used the piecewise Einstein metric \eqref{eq:piecewise}, by first pre-training the neural networks to reproduce them.

An intermediate step, before both scaling up the dimensionalty and including the matter and warping effects, would be the construction of compactifications on three-dimensional spaces, similarly to the case originally studied in \cite[Sec.~8]{deluca-silverstein-torroba}. The method and the code developed and released with the present work should be directly applicable to this problem.

\section*{Acknowledgements}
I would like to thank Eva Silverstein and Gonzalo Torroba for various discussions on the topic of this work and collaboration on related works. In addition, I am grateful to Alice Gatti, Alessandro Tomasiello, and Henry Zheng for comments on an earlier draft of this manuscript. I would also like to thank Lara Anderson, Gauri Batra, Mathis Gerdes, Paolo Glorioso, Jim Halverson, Sven Krippendorf, Albert Law, Fabian Ruehle, Jorge Santos, and Sungyeon Yang for useful discussions. My work is supported in part by the NSF Grant PHY-231042. Some of the computing for this project was performed on the Sherlock cluster. I would like to thank Stanford University and the Stanford Research Computing Center for providing computational resources and support that contributed to these research results.

\appendix
\section{Additional details on the Machine Learning method}\label{app:ML}

For each of the neural networks $\mathcal{N}_i$ entering the metric ansatz \eqref{eq:metNN}, we chose a residual architecture:
\begin{equation}\label{eq:res}
	\mathcal{N}(x;\theta) \equiv x_{D+1} \qquad \text{with}\quad\left\{\begin{array}{lllr}
		x_0 &= W_0 x + b_0 & &\\
		x_i &= x_i+ \sigma(W_i \mathcal{L}_i(x_{i-1}) +b_i ) &\hfill& i = 1, \dots, D\\
		x_D & = W_{D+1}  \mathcal{L}_{D+1}(x_D) + b_{D+1} &&
	\end{array}\right.\:.
\end{equation}
Here $x \in \RR^n$ is the input (with $n = 3$ for the problem studied in this work), $W_0 \in \RR^{n\times H}$, $b_{i = 0,\dots, D}\in \RR^H$, $W_{i= 1, \dots, D} \in \RR^{H \times  H}$, $W_{D+1} \in \R^{H\times 1}$, and $b_{D+1} \in \RR$. 
Each \emph{layer norm} block  $\mathcal{L}_i$ depends on two learnable parameters $\{l_1^i, l_2^i\}$ and is defined as
\begin{equation}
	\mathcal{L}_i(x) \equiv l_1^i \frac{x - \mu}{\sqrt{\sigma^2 + \epsilon}} +l_2^i
\end{equation}
where $\mu$ and $\sigma$ are respectively the mean and variance of $x$, and $\epsilon = 10^{-5}$.  
The collection of all the $W$, $b$, and $l$ forms the set of learnable parameters of the neural network, collectively denoted as $\theta$.  $\sigma$ represents a non-linear function, which is applied element-wise. To ensure differentiability of the network, we opted for the popular choice \mbox{$\sigma = \tanh$}.

$H$ and $D$, respectively the \emph{width} (or \emph{hidden dimension}) and the \emph{depth} of the neural network, control the complexity of the learnable functions. 
We found that using small neural networks \eqref{eq:res}  with $H = 10$, $D = 2$ ($1986$ trainable parameters) is enough to solve the three-dimensional Einstein equations studied in this work.

We also experimented with the improvements suggested in \cite{wang2023expert}, specifically with the \emph{modified MLP} (introduced in \cite{mod-MLP}) and with the \emph{Fourier features} (introduced in \cite{tancik2020fourier}), but we did not find any appreciable improvement on this three-dimensional problem, and thus we opted for the simpler architecture \eqref{eq:res}.

For the pre-training, we used $256 \times 8$ bulk points, changed at each itearation, and to reduce computational cost we only enforced the explicit matching of derivatives on $50\%$ of them and the explicit matching of second derivatives on  $30\%$ of them. We early-stopped the pre-training once the pre-training loss dropped below $3\times 10^{-3}$.

For the training phase, we set the balancing coefficients in \eqref{eq:loss} as $\gamma_{\mathsf{E}} = \gamma_{\mathsf{B}} = 1$ and at each iteration we draw of $16\times 8$ points from the bulk and $32$ points from each element of each of the $14$ pairs of identified boundary components. We used the importance sampling scheme described in Sec.~\ref{sub:ML3d}, retaining $60\%$ of the points with highest error every $10$ steps. We performed multiple training runs both with the optimizer $\mathsf{ECDSep}$ with hyperparameters $\text{lr} = \{1.5, 1\}$, $\nu = 10^{-3}$, $\eta = \{1, 1.5, 2\}$, $F_0 = 0$  and Adam, with hyperparameters $\text{lr} = \{0.001, 0.003, 0.008, 0.01\}$, $\beta_1 = 0.9$, $\beta_2 = 0.999$, $\epsilon = 10^{-8}$, and setting weight decay to zero for both optimizers. We briefly experimented with much largher or much smaller values of the hyperparameters, finding worse performance than in this range. We always started from the same fixed pre-trained network during the hyperparameter exploration.

\section{Additional numerical results}\label{app:detML}

\begin{figure}[ht!]
		\centering
		\begin{subfigure}
			\centering
			\includegraphics[width=.9\linewidth]{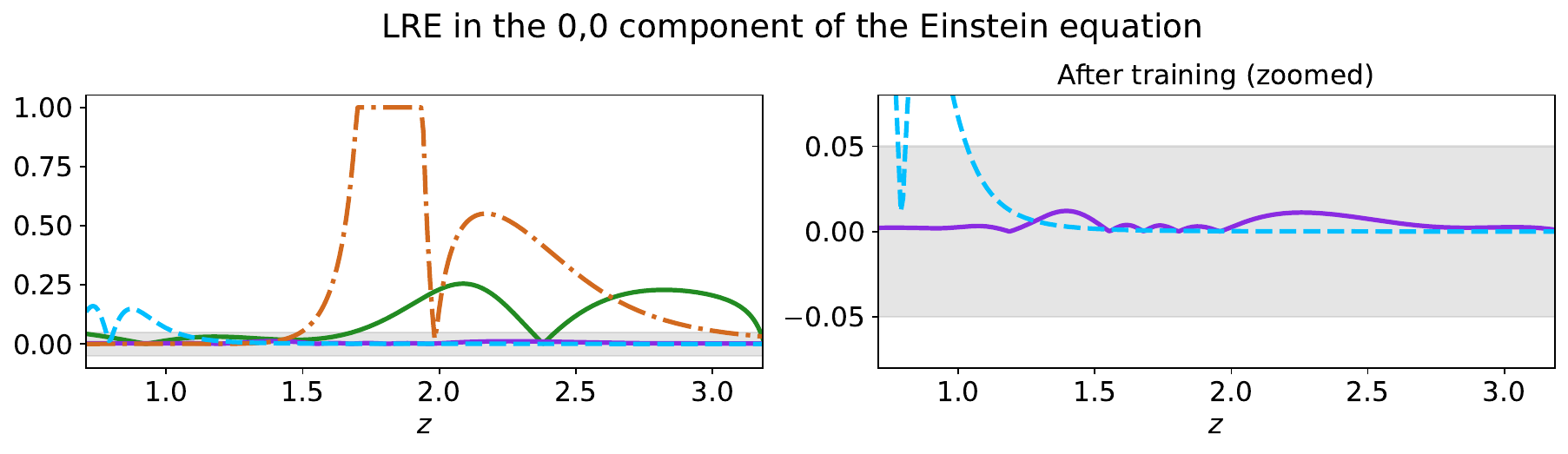}
		\end{subfigure}
		\begin{subfigure}
			\centering
			\includegraphics[width=.9\linewidth]{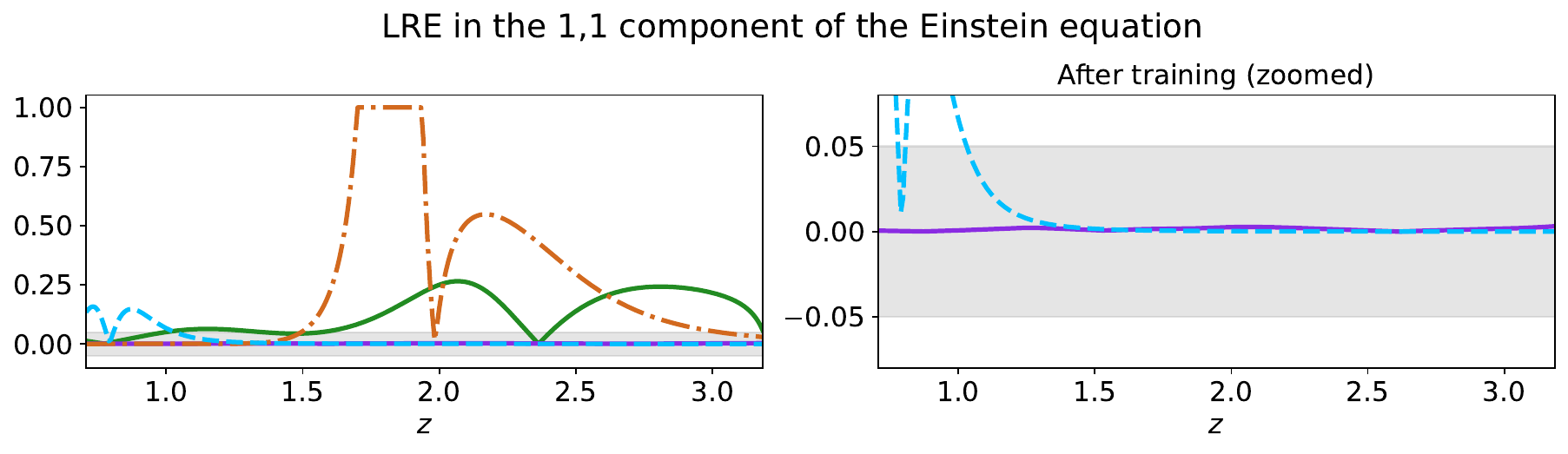}
		\end{subfigure}
		\begin{subfigure}
			\centering
			\includegraphics[width=.9\linewidth]{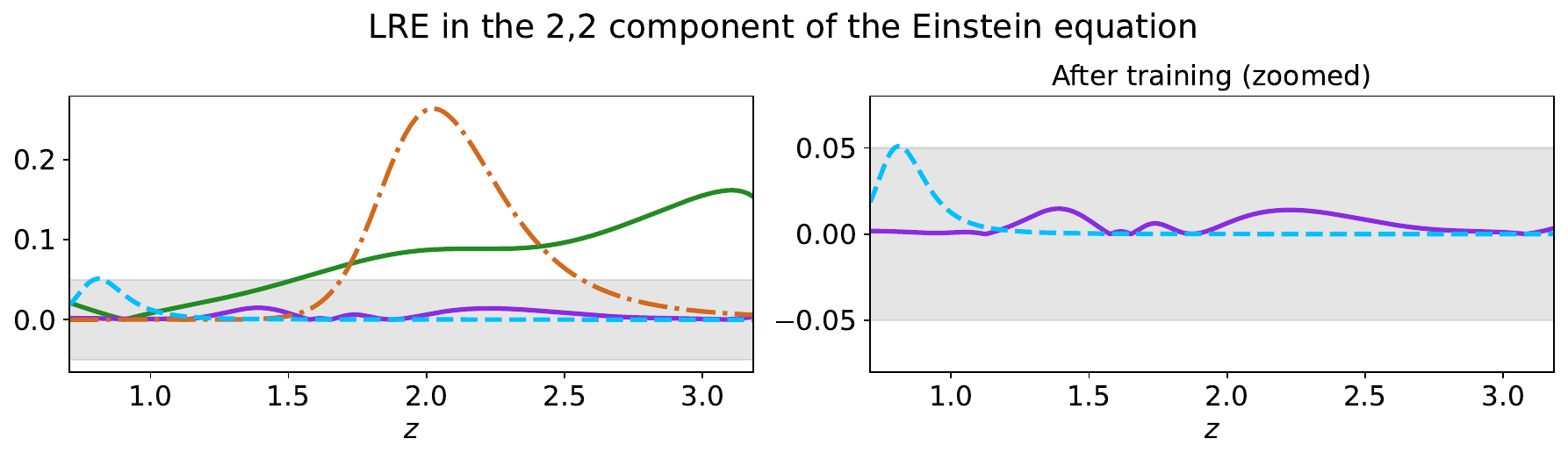}
		\end{subfigure}
		\begin{subfigure}
			\centering
			\includegraphics[width=.9\linewidth]{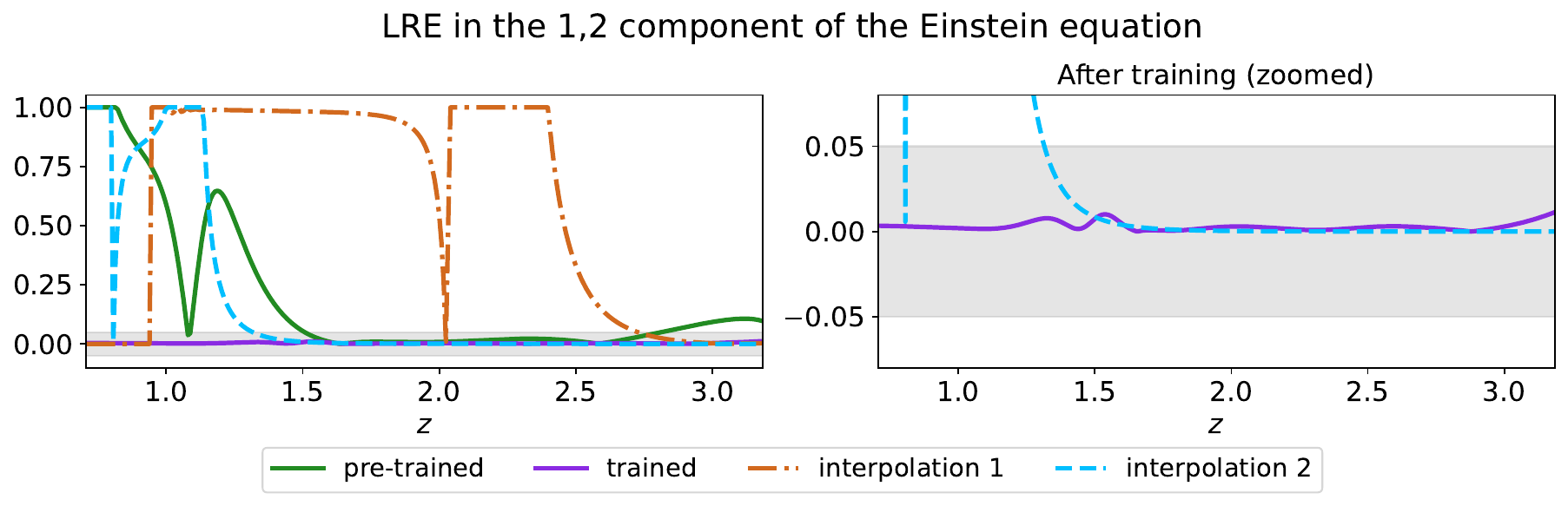}
		\end{subfigure}
	
		\caption{\small Local Relative Error for all the components of the Einstein equation on a typical training run. Shaded area shows 5\% error.}
		\label{fig:ricci_comparison}
	\end{figure}

A family of smooth interpolations of \eqref{eq:piecewise} can be obtained from the metric
\begin{equation}\label{eq:interp}
	 \begin{split}
		ds_\text{interp.} \equiv&\rj^2\frac{dz^2}{z^4\tilde{W}(z)}+         \frac{1}{k^2+\alpha^2}\left[ \left(k^2 \frac{\tilde{W}(z)}{\tilde{W}(1)}+\frac{\alpha^2}{z^2}\right) \dd x_1^2+ \right.                                                                     \\
				&\left. +\left(\alpha^2 \frac{\tilde{W}(z)}{\tilde{W}(1)}+\frac{k^2}{z^2} \right)\dd x_2^2
				 +2 k \alpha\left( \frac{\tilde{W}(z)}{\tilde{W}(1)}- \frac{1}{z^2}\right) \dd x_1 \dd x_2\right]
  \end{split}
\end{equation}
where, following \cite{anderson-dehn} we define $\tilde{W}(z) \equiv z^{-2}\rj^2 -\chi(\frac{\rj}{z})$,  with $\chi(y) \sim 1$ as $y \to \rj/2$ from above and  $\chi(y) \sim 0$ as $r \to 2 \rj$ from below. In addition, to keep the error on the Einstein equations under control, we require $|\partial_y^k \chi| = O(\rj^k)$. A possible choice is 
\begin{equation}
	\chi(y) \equiv \frac12 \left( \tanh\left(- \frac{2 \gamma}{\rj}\left(y -\left(\frac{5}{4} + s\right) \rj\right)\right)+1\right)  \;.
\end{equation}
where larger $\gamma$ results in a faster interpolation, at the expense of larger gradients, and thus of a larger deviation from an Einstein metric, but localized in a smaller region. The interpolation as defined in \cite{anderson-dehn}  is with $s = 0$. However, at $z = 1$ this results in a still large deviation from the hyperbolic metric, resulting in a large error in the boundary conditions, in particular in the extrinsic curvature at the hemispheres. The case $s \neq 0$ below, instead, has a sharper transition to the hyperbolic metric close to the hemispheres, at the price of a larger deviation from the Einstein condition. In our comparisons in Sec.~\ref{sec:res}, we used the two interpolations
\begin{equation}\label{eq:interpolations}
	\begin{array}{ll}
	\text{Interpolation } 1 &\equiv \eqref{eq:interp} \text{ with } \gamma = 8\;, s = -\frac{3}{4} \;, \\
	\text{Interpolation } 2 &\equiv \eqref{eq:interp} \text{ with } \gamma = 3\;, s = 0 \;.	\end{array}
	\end{equation}
In Fig.~\ref{fig:ricci_comparison} we compare the Local Relative Error on all the components of the Einstein equation.

\section{Additional details on $M_3$}\label{app:detM3}

To obtain a smooth manifold at the gluing points we have to correctly impose the boundary conditions on $P_3^8$. To do so in our numerical method, we need to be able to generate pair of points on all the boundaries that have to be identied, according to the procedure described in Sec.~\ref{sub:M3}. 

To generate these pair of points we proceed as follows.
\begin{enumerate}
	\item For each $R_i$, we generate all the extra reflections generated by the new added facets $R_i^{a}$, where $a$ denotes an internal face generated by a reflection $R_i$.
	      Explicitly, the only ones that we need are
	      \begin{equation}
		      \begin{aligned}
			      R_1^a : & R_1^1 = R_1\;,  & R_1^2: x_1\to 4L-x_1\;, \\
			      R_2^a : & R_2^1 = R_2 \;, &                      \\
			      R_3^a : & R_3^1 = R_3 \;. &
		      \end{aligned}
	      \end{equation}
	\item We then generate points on the 6 facetes of $P_3$. For each of these points we generate its 8 copies by acting with $R_1$, $R_2$, $R_3$, as we do for the interior points. Some of these generated points will be duplicates.
	\item Starting from a point $p$, the generated 8 points $p_k$ have to be pairwise identified. To do so, we check whether a given $p_k$ can be obtained from another the $p_{k\prime}$ by acting on it with any of the reflections $R_c^a$ corresponding to a face of the same color as $p$. If yes, $p_k$ and $p_{k\prime}$ have to be identified and we pair them together to impose the boundary conditions.
	\item This procedure produces 4 pairs for each starting $p$. In particular, points that are on ``internal facets'' of $P_3^8$ will turn out to be identified with themselves. We discard such pairs.
\end{enumerate}

This procedure results in 14 pairs of identified faces $(A_i, B_i)$, with $i = 1, \dots, 14$.
We keep track of the hypersurface they belong to in Table \ref{tab:external_faces}. Notice that in this table we are only keeping track of the hypersuface in which the pairs of identified points lie, without keeping track of the boundaries of these hypersurfaces, since we only use this table to construct the corresponding normals and projectors.
\begin{table}[!ht]
	\centering
	\begin{tabular}{c l|c l}
		$A_1$ :                            & Hemisphere centered at $(0, 1, 0)$ & $B_1$ :                             & Hemisphere centered at $(0,- 1, 0)$ \\
		$A_2$ :                            &
		Hemisphere centered at $(0, 1, 2)$ & $B_2$ :                            & Hemisphere centered at $(0,- 1, 2)$                                       \\
		$A_3$ :                            & Hemisphere centered at $(0, 1, 0)$ & $B_3$ :                             & Hemisphere centered at $(0, 3, 0)$  \\
		$A_4$ :                            & Hemisphere centered at $(0, 1, 2)$ & $B_4$ :                             & Hemisphere centered at $(0, 3, 2)$  \\
		$A_5$ :                            & Vertical face at $x_2 = 0$         & $B_5$ :                             & Vertical face at $x_2 = 2$          \\
		$A_6$ :                            & Vertical face at $x_2 = 0$         & $B_6$ :                             & Vertical face at $x_2 = 2$          \\
		$A_7$ :                            & Vertical face at $x_2 = 0$         & $B_7$ :                             & Vertical face at $x_2 = 2$          \\
		$A_8$ :                            & Vertical face at $x_2 = 0$         & $B_8$ :                             & Vertical face at $x_2 = 2$          \\
		$A_9$ :                            & Vertical face at $x_1 = -1$        & $B_9$ :                             & Vertical face at $x_1 = 3$          \\
		$A_{10}$ :                         & Vertical face at $x_1 = 3$         & $B_{10}$ :                          & Vertical face at $x_1 = -1$         \\
		$A_{11}$ :                         & Hemisphere centered at $(0, 0, 1)$ & $B_{11}$ :                          & Hemisphere centered at $(0,2, 1)$   \\
		$A_{12}$ :                         & Hemisphere centered at $(0, 0, 1)$ & $B_{12}$ :                          & Hemisphere centered at $(0, 2, 1)$  \\
		$A_{13}$ :                         & Hemisphere centered at $(0, 0, 1)$ & $B_{13}$ :                          & Hemisphere centered at $(0, 2, 1)$  \\
		$A_{14}$ :                         & Hemisphere centered at $(0, 0, 1)$ & $B_{14}$ :                          & Hemisphere centered at $(0, 2, 1)$  \\\end{tabular}
	\caption{External facets of $P_3^8$ for $L = 1$.}
	\label{tab:external_faces}
\end{table}

\subsection{Maps and Jacobians}
 Consider a vertical hyperplane, with intrinsic coordinates $(\sigma_0, \sigma_1)$. This has to be identified with two vertical facets at either fixed $x_1$ or $x_2$, as in Table \ref{tab:external_faces}.
For these facets, the projectors defined in Sec.~\ref{sub:bry} are position independent. In particular, for the vertical faces at  fixed $x_1$ we have
\begin{equation}
	e^m_a =    \begin{pmatrix}
		1 & 0 & 0 \\
		0 & 0 & 1
	\end{pmatrix}\;,\qquad\qquad \text{fixed } x_1\;,
\end{equation}
while for the ones at fixed $x_2$ we have
\begin{equation}
	e^m_a =    \begin{pmatrix}
		1 & 0 & 0 \\
		0 & 1 & 0
	\end{pmatrix}\;,\qquad\qquad \text{fixed } x_2\;.
\end{equation}

For the hemispheres, instead the embedding maps are:
\begin{equation}
	\alpha_i : \begin{cases}
		z = \sqrt{1-\sigma_1^2-\sigma_2^2} \\
		x = c_1+\sigma_1                   \\
		y = c_2+\sigma_2
	\end{cases}\;,
	\label{eq:alphai}
\end{equation}
and
\begin{equation}
	\beta_i : \begin{cases}
		\hat{z} = \sqrt{1-\sigma_1^2-\sigma_2^2} \\
		\hat{x} = \hat{c}_1-\sigma_1             \\
		\hat{y} = \hat{c}_2+\sigma_2
	\end{cases}
	\label{eq:betai}\;,
\end{equation}
where $\hat{c}$ denotes the center of the hemisphere $B_i$.
Thus, the Jacobian $e_A$ is
\begin{equation}
	e_A (\sigma) = \left(
	\begin{array}{ccc}
		-\frac{\sigma_1}{\sqrt{1-\sigma_1^2-\sigma_2^2}} & 1 & 0 \\
		-\frac{\sigma_2}{\sqrt{1-\sigma_1^2-\sigma_2^2}} & 0 & 1 \\
	\end{array}
	\right)\;,
\end{equation}
while $e_B$ can be compute by acting on it with $J$ defined in Sec.~\ref{sub:bry}.

Summarizing, to impose continuity we have to impose in the code \eqref{eq:hab}, where
$x(\sigma)$ and $\hat{x}(\sigma)$ are the point connected by the reflection, and we write $e_A(\sigma)$ as function of $x$ as
\begin{equation}
	e_A(\sigma) = e_A (x(\sigma)) =  \left(
	\begin{array}{ccc}
		\frac{c_1-x_1}{z} & 1 & 0 \\
		\frac{c_2-x_2}{z} & 0 & 1 \\
	\end{array}
	\right)\;.
\end{equation}
$e_B$ is then obtained by acting with $J$ on $e_A(\sigma)$.

The same quantities are needed to impose differentiability using \eqref{eq:Kab}.

\bibliography{refs}
\bibliographystyle{at}

\end{document}